\documentclass{aa}
\usepackage{graphicx}
\usepackage{txfonts}

\usepackage{hyperref}
\usepackage{diagbox}
\usepackage{rotating}
\usepackage{tikz}
\usepackage{multirow}
\usepackage{wasysym}
\usepackage{adjustbox}
\usepackage{afterpage}
\usepackage{placeins}
\usepackage{xcolor}
\usepackage[strict]{changepage}
\usepackage{lscape}
\usepackage{pdflscape}
\usepackage{soul}

\hypersetup{colorlinks=True,citecolor=blue}

\begin{document}

\def\ms{\hbox{\,m\,s$^{-1}$}}         
\def\cms{\hbox{\,cm\,s$^{-1}$}}       
\def\m2s2{\hbox{\,m$^{2}$\,s$^{-2}$}} 
\def\kms{\hbox{\,km\,s$^{-1}$}}       
\def\vsini{\hbox{$v$\,sin\,$i$}}      
\def\sini{\hbox{sin\,$i$}}      
\def\Msun{\hbox{$\mathrm{M}_{\odot}$}}             
\def\Mearth{\hbox{$\mathrm{M}_{\oplus}$}}             
\def\Rearth{\hbox{$\mathrm{R}_{\oplus}$}}             
\def\Rsun{\hbox{$\mathrm{R}_{\odot}$}}
\def\Mjup{\hbox{$\mathrm{M}_{\rm Jup}$}}
\def\Rjup{\hbox{$\mathrm{R}_{\rm Jup}$}}
\def\degr{\hbox{$^\circ$}}
\def\chisq{\mbox{$\chi^2$}}
\def\mp{M_{\rm p}}
\def\rp{R_{\rm p}}
\def\logrhk{$\log$(R$^{\prime}_{HK}$)}

\newcommand{\xavier}[1]{{\color{blue}[[\textbf{Xavier: }#1]]}}
\newcommand{\LEt}[1]{{\color{red}[[\textbf{Editor: }#1]]}}
\newcommand{\michael}[1]{{\color{red}[[\textbf{Michael: }#1]]}}

   \title{Measuring precise radial velocities on individual spectral lines}
   \subtitle{II. Dependance of stellar activity signal on line depth}

   \author{M. Cretignier\inst{1}
          \and X. Dumusque \inst{1}
          \and R. Allart \inst{1}
          \and F. Pepe \inst{1}
          \and C. Lovis \inst{1}
          }

   \institute{Astronomy Department of the University of Geneva, 51 ch. des Maillettes, 1290 Versoix, Switzerland\\
              \email{michael.cretignier@etu.unige.ch}
 }

   \date{Received XXX ; accepted XXX}

  \abstract
   {Although the new generation of radial-velocity (RV) instruments such as ESPRESSO are expected to reach the long-term precision required to find other earths, the RV measurements 
   are contaminated by some signal from stellar activity. This makes these detections hard.}
   {Based on real observations, we here demonstrate for the first time the effect of stellar activity on the RV of individual spectral lines. Recent studies have shown that this is probably the key for mitigating this perturbing signal.}   
   {By measuring the line-by-line RV of each individual spectral line in the 2010 HARPS  RV measurements of $\alpha$ Cen B, we study their sensitivity to telluric line contamination and line profile asymmetry. After selecting lines on which we are confident to measure a real Doppler-shift, we study the different effects of the RV signal that is induced by stellar activity on spectral lines based on their physical properties.}
   {We estimate that at least $89\%$ of the lines that appear in the spectrum of $\alpha$ Cen B for which we measure a reliable RV are correlated with the stellar activity signal (Pearson correlation coefficient $\mathcal{R}>0.3$ at $2\sigma$). This can be 
   interpreted as those lines being sensitive to the inhibition of the convective blueshift observed in active regions. Because the velocity of the convective blueshift increases with physical depth inside the stellar atmosphere, we find that the effect induced by stellar activity on the RV of individual spectral lines is inversely proportional to the line depth. The stellar activity signal can be mitigated down to $\sim$ 0.8-0.9 \ms either by selecting lines that are less sensitive to activity or by using the difference between the RV of the spectral lines that are formed at different depths in the stellar atmosphere as an activity proxy.
   }
   {This paper shows for the first time that based on real observations of solar-type stars, it is possible to measure the RV effect of stellar activity on the RV of individual spectral lines. 
  Our results are very promising and demonstrate that analysing the RV of individual spectral lines is probably one of the solutions to mitigate stellar activity signal in RV 
  measurements down to a level enabling the detection of other earths.}


   \maketitle
%
\section{Introduction}\label{introduction}

To characterise exoplanets in depth, it is crucial to measure their radius and mass, which gives their density. This information is critical for understanding the core composition of exoplanets and thus how they formed, but also for properly interpreting signatures in their atmospheres (e.g. \citet{Ehrenreich(2006), Lecavelier(2008), Kreidberg(2014), Heng:2017aa}). The CoRoT, Kepler, and TESS satellites have measured the radius of several thousand exoplanets, but fewer than one thousand of them have a determined mass\footnote{https://exoplanetarchive.ipac.caltech.edu/index.html}. When we restrict ourselves to exoplanets with a radius smaller than 1.5\,\Rearth\, that orbit stars brighter than magnitude 12, for which the RV method can obtain a sufficient precision to measure exoplanet masses, only one-fourth out of the 73 discoverd exoplanets have an RV-measured mass. This small number highlights that it is difficult to measure the mass of small exoplanets with the RV method, mainly because of the perturbations induced by stellar signals. 

The most significant of the different types of stellar signal (e.g \citet{Meunier(2017), Dumusque(2011e), Meunier(2010), Lindegren-2003}) and the signal that is most difficult to work with is the activity signal that is induced by magnetic regions that rotate with the stellar surface (e.g \citet{Saar-1997b, Meunier(2010), Dumusque(2014)}). In the case of an Earth analogue, the signal induced by the exoplanet is more than a magnitude smaller than the activity signal (e.g \citet{Dumusque:2017aa, Dumusque-2016a, Borgniet(2015), Meunier(2010)}). Therefore, even though the new generation of RV instruments such as ESPRESSO \citep{Pepe(2014),Gonzales(2017)}, EXPRESS \citep{Fischer:2017aa}, and NEID \citep{Schwab:2019aa} are designed to reach the 10\,\cms\,precision that is required to detect and measure the mass of Earth analogues, it is critical to understand and mitigate the signal induced by stellar activity.

Several techniques have been developed to mitigate the stellar activity signal in RV measurements (e.g. \citet{Dumusque:2017aa}). It is possible to de-correlate the stellar activity signal in RVs using proxies that probe magnetic regions on the stellar surface, such as the \logrhk, activity index \citep{Wilson-1968, Noyes-1984}, H$\alpha$ (e.g \citet{Robertson-2014}), and others (e.g. \citet{West:2008aa, Maldonado:2019aa}), or by measuring line profile variations of the cross-correlation function (CCF) using the bisector span and its variant (BIS SPAN, \citet{Queloz-2001, Boisse-2011, Figueira-2013, Simola:2018aa}), and the full width at half-maximum (FWHM) of the Gaussian fitted CCF (e.g. \citet{Queloz-2009}). Another possibility to mitigate stellar activity is to model its correlation on the stellar rotation timescale using a Gaussian process, moving average, or kernel regression technique (e.g \citet{Tuomi-2013, Haywood(2014), Rajpaul(2015), Jones(2017), Lanza:2018aa}). Although all these mitigation techniques have been used in several publications with more or less success, they all fail in the detection of Earth analogues using the RV technique \citep{Dumusque:2017aa}.

All the techniques presented above to mitigate stellar signals used either information coming from the CCF (RV, BIS SPAN, and FWHM) or from chromospheric indicators.
\citet{Davis(2017)} was the first to show that different spectral lines exhibit different behaviour when
they are affected by stellar activity. Then, \citet{Thompson(2017)} and \citet{Wise(2018)} showed more precisely that the depth and equivalent width (EW) of some spectral lines were strongly correlated to stellar activity. However, a change in depth or EW does not necessarily imply a change in RV because these variations do not create an asymmetry of the spectral line profile. \citet{Dumusque(2018)} measured the RV of individual spectral lines and showed that the RVs of some lines are much more sensitive to stellar activity than others, which allowed the author to mitigate the stellar activity signal by a factor two by carefully choosing the spectral lines on which the RV is measured. However, this implies that some lines are not used when the RV is estimated, which increases the photon noise. In addition, the proposed technique requires measuring the RV of individual spectral lines, which can only be done for a significant number of spectral lines for very bright stars. To be able to apply this technique to other stars, we must understand the physics behind the interaction of a spectral line and the magnetic field in active regions.

 \citet{Reiners(2016)} and \citet{Dravins-1981} showed that the convective blueshift of spectral lines depends on their depth. For slow rotators, the inhibition of convective blueshift in active regions is the dominant effect of stellar activity (e.g \citet{Meunier(2010)}), we therefore expect spectral lines of different depths to be affected differently by stellar activity,  as was shown in the simulation published in \citet{Meunier:2017aa}. The authors meant to measure this effect on real observations, but concluded that their data were not good enough.

The goal of the current paper is to improve the precision at which we are able to measure the RV of individual spectral lines and try to measure the differential effect of stellar activity on spectral lines of different depths. The paper is organised as follow. In Sect.~\ref{mean_RV} we discuss the data that we used in this analysis. In Sect.~\ref{line_by_line_RV} we improve the way of computing the RV of individual spectral lines with regard to \citet{Dumusque(2018)}. We also investigate the RV effect induced by the profile of spectral lines and by telluric contamination. After selecting spectral lines of which we are confident that the measured RV effect is real, we are able to show in Sect.~\ref{CB_inhibition} that spectral lines of different depths are affected differently by stellar activity. We also find that the RV effect induced by stellar activity is inversely proportional to the line depth. We finally discuss different techniques that might be used to mitigate stellar activity following this new result, and we conclude in Sect.~\ref{conclu}.

\section{Radial velocities for $\alpha$ Cen B} \label{mean_RV}

\begin{figure}[tp]
        \centering
        \includegraphics[width=10cm]{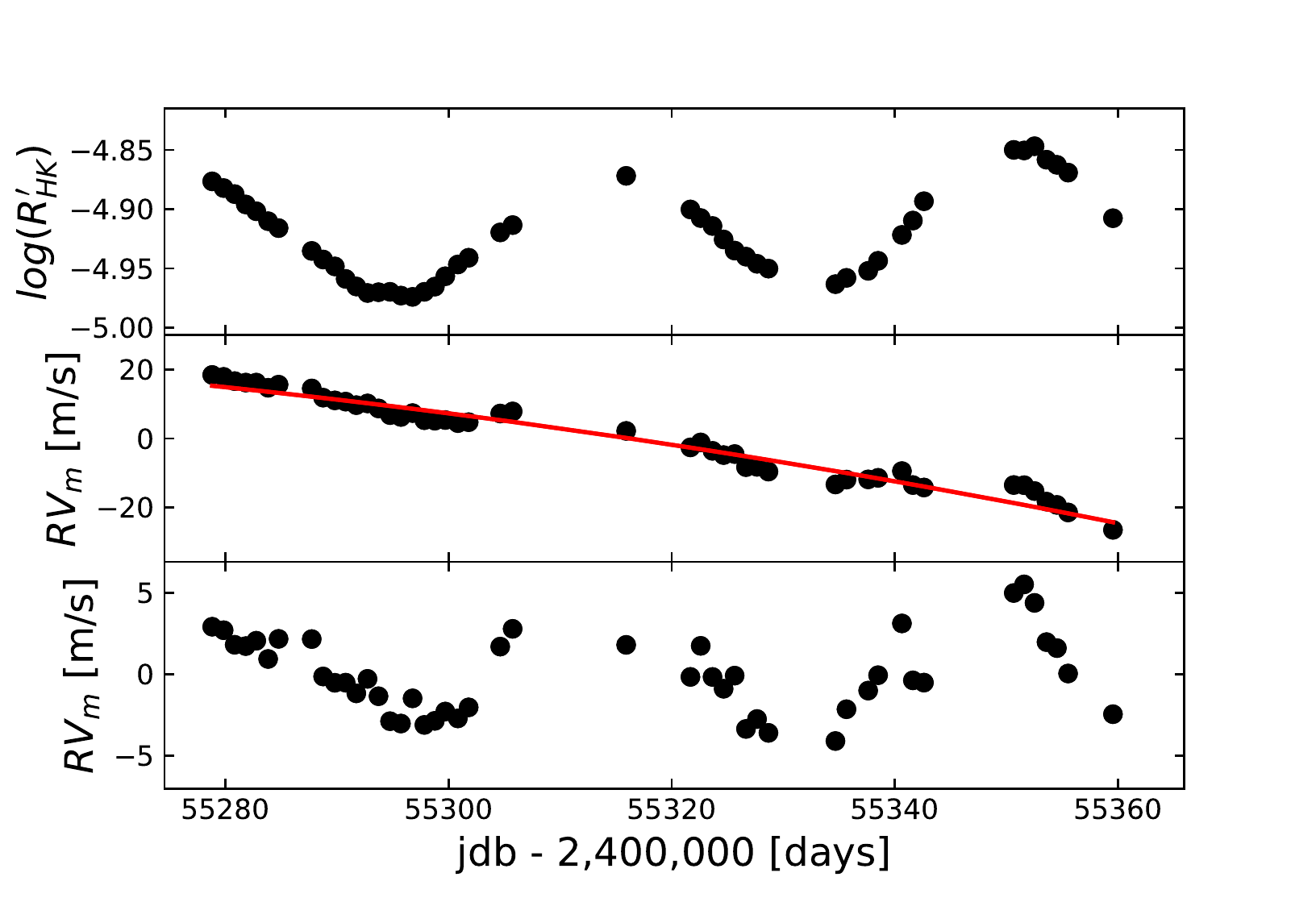}
        \caption{\textbf{Top}: Calcium activity index \logrhk\,time series of $\alpha$ Cen B in 2010. The observations span two consecutive rotational phases, where $P_{rot}=35.8$ days. \textbf{Middle}: RV time series over the same time interval. We show a quadratic trend fitted to the data (red) that corresponds to the binary contribution of $\alpha$ Cen A and a possible activity cycle \citep[][]{Dumusque(2012)}. \textbf{Bottom}: RV residuals after removing the quadratic trend, which present an RV rms of 2.1\ms. These RV residuals are strongly correlated to the calcium activity index and therefore are significantly affected by stellar activity.}
        \label{FigRvRHK}
\end{figure}

We analysed the 2010 RV measurements of $\alpha$ Cen B, which are  composed of 47 nights of observation over a total span of 81 days
\citep{Dumusque(2012)}. This data set has been intensively used in previous studies to analyse the effect of stellar activity on RV measurements (e.g \citet{Dumusque(2018), Wise(2018), Thompson(2017), Dumusque(2014b)}). During these 81 days,
the \logrhk\,activity index shows a sinusoidal variation with a period of 35.8 days (see top panel of Fig.~\ref{FigRvRHK}).
This period is close to the 37-day stellar rotation period derived from other studies \citep{Jay(1997),Dewarf(2010)}. The peak-to-peak amplitude of this activity signal is equal to 0.15 dex, which for a quiet star like $\alpha$ Cen B, is a wide range to exhibit over a stellar rotation timescale. By comparison, this peak-to-peak amplitude is displayed by the Sun beween the minimum and maximum phase of the solar cycle.

In the RVs of $\alpha$ Cen B, shown in the middle panel of Fig.~\ref{FigRvRHK}, a long-term trend is induced by  the binary companion $\alpha$ Cen A and by a possible activity cycle \citep[][]{Dumusque(2012)}. In addition, a sinusoidal variation around this long-term trend is produced by the stellar activity on a rotational timescale that is correlated with the activity index. To prevent biasing the fit of the long-term trend with the sinusoidal signal present in the data, we fitted the RVs of $\alpha$ Cen B with a quadratic polynomial (red line in the middle panel of Fig.~\ref{FigRvRHK}) in addition to a sinusoidal signal, whose period was allowed to vary in the fit. This converged to a period of 35.8 days, identical to the one found in \logrhk. After removing the quadratic trend from the RVs, the RV residuals shown in the bottom panel of the same figure were left, which are strongly correlated to the \logrhk.
This strong correlation suggest that the 2010 RV measurements of $\alpha$ Cen B are strongly contaminated by stellar activity, with a signal exhibiting an RV rms of 2.1\ms.

\begin{figure*}[h]
        \centering
        \includegraphics[width=18.7cm]{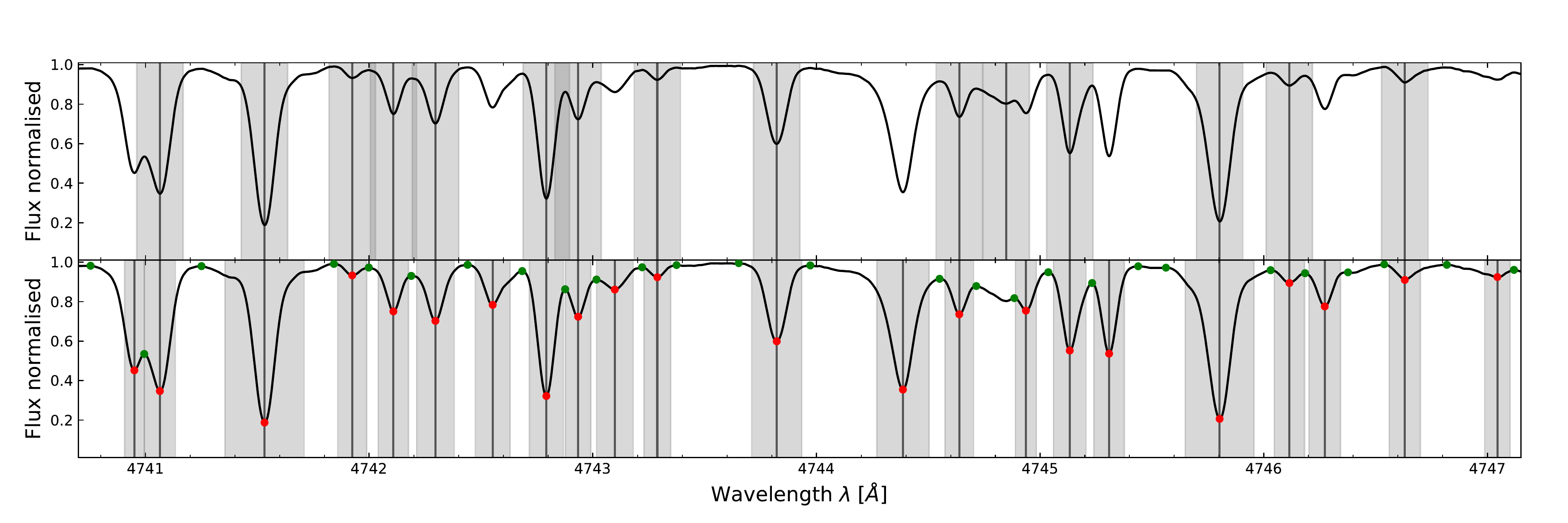}
        \caption{Comparison between the line selection used in \citet{Dumusque(2018)} and the new line selection described in this work. \textbf{Top panel}: Line centres given by the HARPS K5 mask and window width fixed at 16 pixels \citep[see][]{Dumusque(2018)}. \textbf{Bottom panel}: Lines are automatically found using an extremum localisation algorithm. The minima shown in red give the line centres, and the maxima shown in green define the maximum width allowed for the windows. Our windows are narrower than these maximum windows and are derived using the second derivative of the flux (see Appendix~\ref{appendixB} for more details).}
        \label{FigMask}
\end{figure*}

With an estimated rotational period of 35.8 days and a stellar radius of 0.863\,\Rsun\,\citep{Kervella-2003}, $\alpha$ Cen B is an extremely slow rotator with $v_{\mathrm{rot,\,equatorial}}$=1.22\,\kms. Based on this low rotational velocity, we can infer from stellar activity simulations that the RV effect induced by the inhibition of convection
inside faculae probably dominates the RV variation induced by the flux effect originating from dark spots \citep[e.g.][]{Dumusque(2014), Meunier(2010)}. This has
been confirmed for the Sun \citep[$v_{\mathrm{rot,\,equatorial}}\sim$2\,\kms,][]{Milbourne(2019), Haywood(2016)}. In addition, \citet{Dumusque(2014b)} showed that the observed RV variation in the 2010 data of $\alpha$ Cen B can be explained by the presence of a single large facule on the stellar surface. Based on the results found in these previous studies, we decided to neglect the RV contribution coming from spots here, although we understand that this might limit this study.

\section{Line-by-line RVs for $\alpha$ Cen B} \label{line_by_line_RV}

To derive the velocity of individual spectral lines, we selected for each of them a centre and a window in which to derive the RV.
In contrast to \citet{Dumusque(2018)}, who used the centre of each line as defined in the K5 cross-correlation mask used on HARPS and a fixed window of 16 pixels around each line centre, 
we performed here a tailored selection for each line based on a spectrum with a high signal-to-noise ratio (S/N) of $\alpha$ Cen B, obtained by stacking nearly 200 HARPS spectra (see Appendix \ref{appendixAA}).

In Fig.~\ref{FigMask} we compare the line selection used in \citet{Dumusque(2018)}, shown in the top panel, with our new selection used hereafter, shown in the lower panel. The figure shows that the selection of the line at 4746.4\,$\AA$ in the top panel is not well centred and its profile within the selected window is clearly asymmetric. In addition, the windows of neighbouring lines sometimes overlap, for example the two lines near 4742.9 $\AA$. In addition to adding some redundancy to the information measured, the RV of such lines will be contaminated by that of their neighbours, which is not desired. A comparison of the upper and lower panels in Fig.~\ref{FigMask} shows that our new line selection, based on the medium window (MW, described in Appendix~\ref{appendixAA}), considers more spectral lines, defines more appropriate windows and more precise centre, and rejects spurious lines.

We obtained the RVs of 7860 out of the 5830 spectral lines in our line selection. This number is larger than the number of lines in the selection because we analysed the
2D HARPS echelle spectra, and because consecutive orders may overlap, the same line can appear twice. To obtain a single RV information for each twin line pair, we performed
a weighted average using as weight the RV precision of the two lines. In the following, we refer to the RV of single lines as $RV_{i}$, in contrast to $RV_{m}$, which corresponds to the weighted average of all $RV_{i}$, using as weight the inverse square of the RV photon-noise error of each line $i$.

\subsection{Removing telluric contaminated lines} \label{remove_tellurics}

\begin{figure}[tp]
        \centering
        \includegraphics[width=9cm]{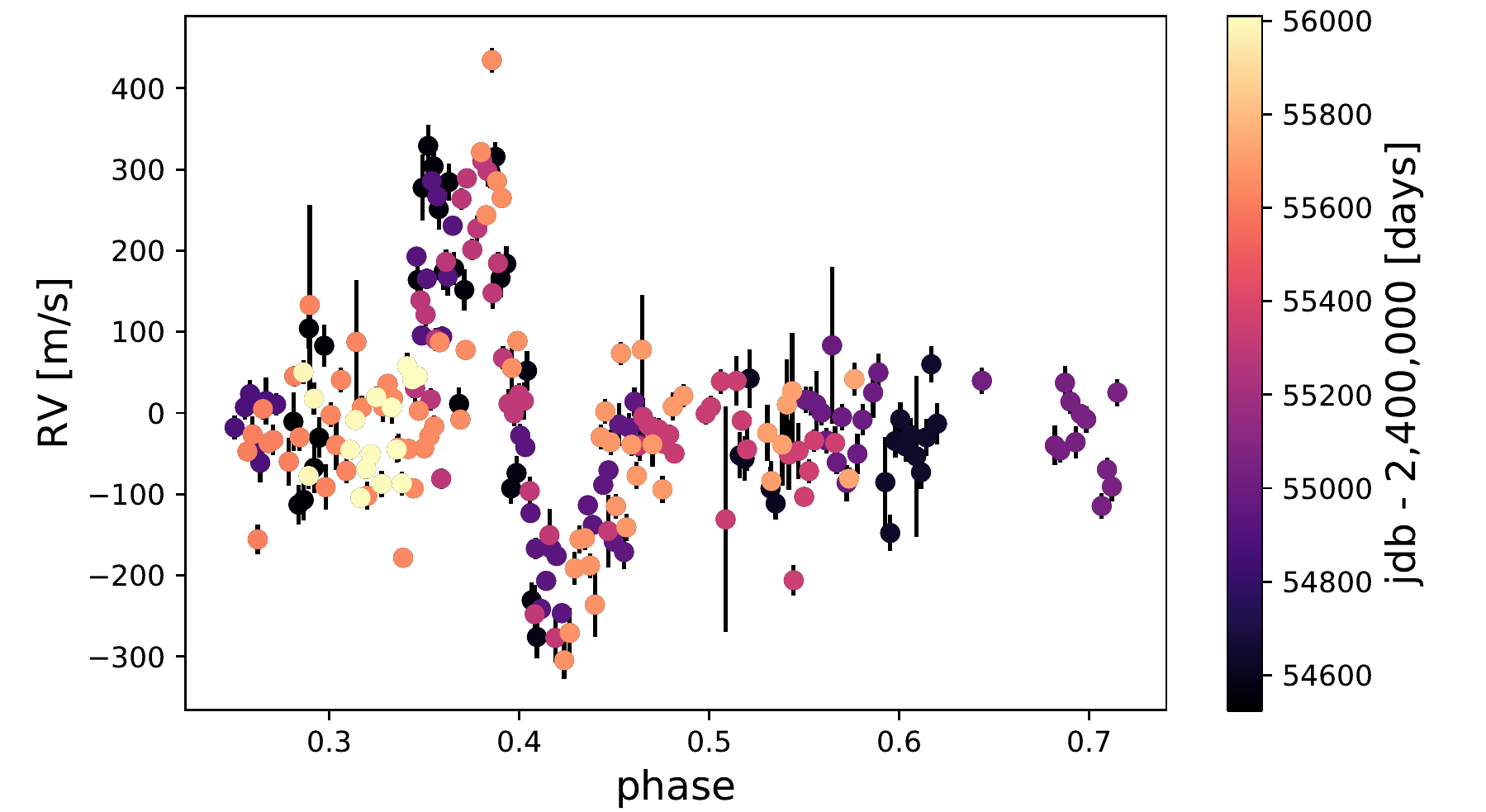}
        \caption{RV$_{i}$ time series of the shallow
        Cr I line at 5038.90\,$\AA$, from 2008 to 2012, phase-folded with the period of a year. This stellar line is contaminated by the 5038.85 $\AA$ telluric line, which has a depth ratio of 5\%. A peak-to-peak RV variation of 600\ms\,is observed when the spectral line moves through the telluric line over the year.}
        \label{FigTel}
\end{figure}

\begin{figure*}[h]
        \centering
        \includegraphics[width=19cm]{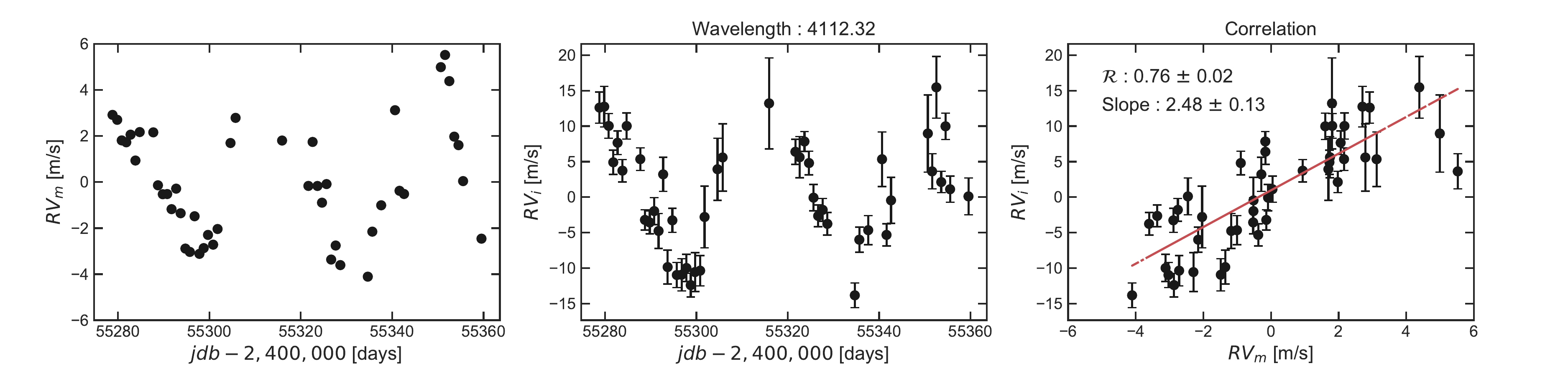}
        \caption{\textbf{Left panel}: Detrended one-day binned mean radial velocity $RV_{m}$ of $\alpha$ Cen B measured on all the spectral lines. Uncertainties are displayed, but are too small to be visible. \textbf{Middle panel}: Detrended one-day binned individual radial velocity $RV_{i}$ of the Fe I line at $4112.30$ \text{\AA}. \textbf{Right panel}: Weighted linear regression between $RV_{i}$ and $RV_{m}$ (\textit{red line}). The coefficient $\mathcal{R}$ as well as the correlation slope are indicated in the legend. The peak-to-peak variation due to activity is $\sim$10\ms\,in the $RV_m$ and $\sim$30\ms\,in the $RV_i$ of the Fe I 
                line at $4112.32$ $ \text{\AA}$.}
        \label{FigCorrelation}
\end{figure*}

Telluric lines contaminate stellar lines and thus affect our derived $RV_i$. Telluric lines are localised and weak in the visible, but they still affect a significant fraction of stellar lines \citep[e.g.][]{Artigau:2014aa, Cunha:2014aa}

As an example, we display in Fig.~\ref{FigTel} the $RV_i$ of the shallow 
Cr I line at 5038.90\,$\AA,$ which has a depth of 0.11 compared to the stellar continuum. This line is contaminated by a telluric line at 5038.85 $\AA$ that has a depth of 0.006, thus a depth ratio of 5\%.
This figure shows that the RV of this line, measured between 2008 and 2012 and phase folded with a period of a year, is strongly contaminated by the 5\% telluric line and shows an RV 
peak-to-peak variation of 600\,\ms. The redshifted and blueshift anomalies coincide in time with the telluric line, which lies on the right and left wing of the stellar line, respectively.
Although the induced RV perturbation is not physically correlated with activity, we decided to reject stellar lines that were contaminated by telluric lines stronger than 2\% in depth ratio to prevent any spurious correlations 
between $RV_i$ and the stellar activity signal (see Sect.~\ref{RV_per_line_group}).

To detect the stellar lines in the 2010 spectra of $\alpha$ cen B that are contaminated by telluric lines, we first modelled a telluric line spectrum by fitting a spectrum with a high S/N of $\alpha$ Cen B using MolecFit \citep{Smette:2015aa, Allart(2017)} and recorded the central wavelengths of the telluric lines, determined by their local minimum position. The spectral windows defined for each line in Sect.~\ref{line_by_line_RV} were obtained from spectra that were taken at a barycentric Earth RV (BERV) of 14.6\,\kms, whereas during the 2010 observations of $\alpha$ Cen B, the BERV changed from 18.8 to $-7.7$\,\kms.  We therefore rejected each stellar line for which its respective window was closer than 4.2 and $-$22.3\,\kms\,to a 2\% telluric line in depth ratio.

After removing contaminated lines from our new selection defined in Sec.~\ref{line_by_line_RV}, we are left with 5753 unique lines stronger than $5\%$, which corresponds to 1.45 times more lines than in the HARPS K5 mask used in \citet{Dumusque(2018)} after the same criterion for telluric contamination was taken into account. This new line selection gives us a non-negligible increase of  9\% in the RV information.

\begin{figure*}[t]
        \centering
        \includegraphics[width=18cm]{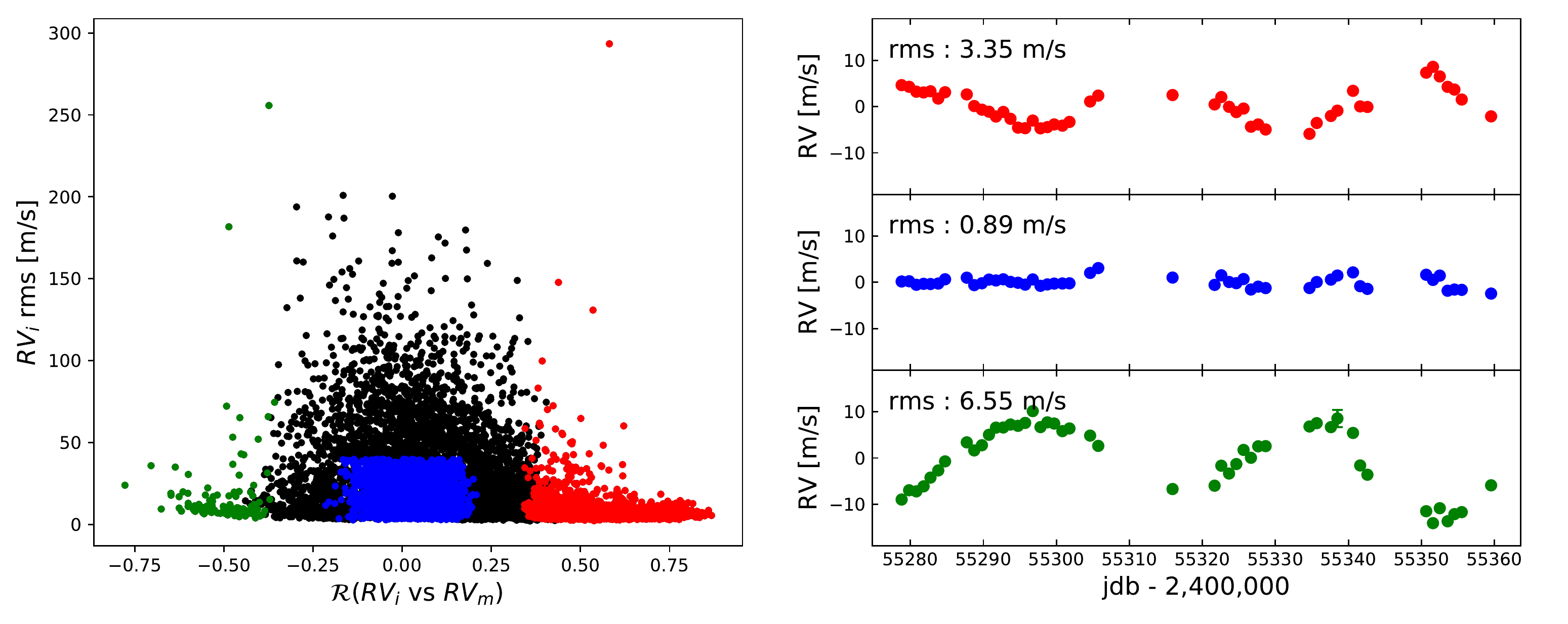}
        \caption{\textbf{Left}: $RV_i$ rms as a function of the Pearson correlation coefficient between $RV_{i}$ and $RV_{m}$ (black points). The different colours correspond to the groups defined in Table~\ref{TableClass}: correlated (red), uncorrelated (blue), and anti-correlated (green). \textbf{Right}: RV time-series obtained by combining the RV information of all the lines in each group defined in Table~\ref{TableClass} and shown in the left panel.}
        \label{Fig3groups}
\end{figure*}

\subsection{RV for different selections of spectral lines} \label{RV_per_line_group}

After an optimal spectral line selection was performed and lines contaminated by telluric lines were removed, we analysed the sensitivity of each spectral line to stellar activity. We measured the Pearson correlation coefficient $\mathcal{R}$ and the slope of the linear
regression between $RV_{i}$ and an activity indicator. We might also have used \logrhk, but because this activity index probes the stellar chromosphere, there is not necessarily a one-to-one correlation with the stellar photosphere. We therefore preferred to use $RV_{m}$ here as the activity indicator because no significant planetary signal was found in the RV data of  $\alpha$ Cen B\footnote{\citet{Dumusque(2012)} published the detection of a planetary candidate with a semi-amplitude of 0.5 m/s. This detection was called into question by \citet{Rajpaul-2016}. Even if the planet exists, the amplitude of its signal is an order of magnitude smaller than the activity signal observed in 2010, therefore we can use $RV_{m}$ as an activity proxy.}

Before we calculated for each line the correlation between $RV_{i}$ and $RV_{m}$,
we first removed from all time series the long-term trend shown in the middle panel of Fig.~\ref{FigRvRHK}. We also removed outliers found in the 
$RV_i$ by computing the distance between the first and thirrd interquartiles (Q1 and Q3) and rejecting any data point that lies farther out from Q1 or Q3 than twice this distance\footnote{We chose this interquartile-clipping rather than a standard sigma-clipping because the standard deviation used in the latter is more sensitive to outliers.} \citep[][]{Upton(1996)}.
To assess the statistical significance of the correlation coefficient $\mathcal{R}$ and the slope, we computed their uncertainty by performing 1000 independent shufflings of the $RV_i$ according to a normal probability distribution with a dispersion equal to the $RV_i$ error bars. 

An example of the correlation between $RV_{i}$ and $RV_{m}$ for the 
Fe I line at 4112.32\,$\AA$ is shown in Fig.~\ref{FigCorrelation}.
The observed strong correlation, with a Pearson correlation coefficient of $\mathcal{R}=0.76 \pm 0.02$, 
suggests that this line is strongly sensitive to stellar activity.

In the left panel of Fig.~\ref{Fig3groups} we show the $RV_i$ rms as a function of the Pearson correlation coefficient between $RV_{i}$ and $RV_{m}$. We can distinguish three categories of spectral lines based on their correlation with stellar activity: correlated, uncorrelated, and anti-correlated, as defined by the criteria given in Table~\ref{TableClass}. In the right panel of the same figure, we show the RV measured by averaging the $RV_{i}$ of each category of lines. The stellar activity signal can be amplified when only the correlated or anti-correlated lines are considered, or it can be mitigated by studying only the uncorrelated lines. The RV on correlated and uncorrelated spectral lines has previously been measured by \citet{Dumusque(2018)}, who showed for the first time that a tailored selection of lines can significantly mitigate stellar activity. The best RV rms obtained by \citet{Dumusque(2018)} for the uncorrelated lines was 1.21\ms, 1.6 times smaller than the rms of the RVs measured using all the spectral lines. Here, our selection of uncorrelated lines is able to mitigate stellar activity down to 0.89\ms, twice smaller than the rms of the original RVs. This improvement was made possible by our new selection of lines, tailored to the spectrum of $\alpha$ Cen B.
\renewcommand{\arraystretch}{1.6}
\begin{table}[tp]
        \footnotesize
        \caption{Criteria to form the groups of lines that are correlated, uncorrelated, and anti-correlated to stellar activity based on two parameters: the Pearson coefficient $\mathcal{R}$ of the correlation between $RV_{i}$ and $RV_{m}$ , and the rms of $RV_i$ in \ms. We note that the inequality on $\mathcal{R}$ has to be satisfied by more than 2$\sigma$. Because high jitter can reduce the correlation of a line with respect to stellar activity, we imposed a threshold of 40 m/s in the $RV_i$ rms for the uncorrelated group. We indicate in the last column the number of lines in each group and the percentage of RV information that it represents compared to the complete line selection.}
        \label{TableClass}
        \centering
        \begin{tabular}{ccc}
                \hline\hline

                Category & Criterion & Number of lines\\
                \hline

                Correlated & $\mathcal{R} > 0.3\,\mathrm{at}\,2\sigma$ & 1619 (54\%) \\
                Uncorrelated & $|\mathcal{R}| < 0.25\,\mathrm{at}\,2\sigma \; \& \;  rms < 40\ms $  & 1871 (15\%)  \\
                Anti-correlated & $\mathcal{R} < -0.3\,\mathrm{at}\,2\sigma$  & 103 (2\%)\\
                \hline
        \end{tabular}
\end{table}

 A correlation can be explained by the fact that because stellar activity inhibits convection inside active regions, the star appears redshifted, which corresponds to a positive RV. Therefore, the more active the star, the higher the RV, hence the positive correlation. With our current knowledge of stellar activity, a physical process that could induce an anti-correlation in solar-type stars is hard to imagine. However, if the origin of the anti-correlated lines were physical, we would expect these spectral lines to have different physical parameters. We therefore cross-matched all the selected lines with VALD 3, and looked for a difference in the physical parameters of the correlated and anti-correlated lines (see Appendix~\ref{appendixA}). Fig.~\ref{FigKDE1} shows that it is difficult to explain correlated and anti-correlated lines using common physical parameters for the lines. We therefore determined whether the morphology of the spectral lines might be at the origin of the behaviour, and indeed, all the anti-correlated lines present strongly asymmetric line profiles. When correlated and anti-correlated lines are compared with respect to the different morphological parameters defined in Appendix~\ref{appendixB}, we clearly see in Fig.~\ref{FigKDE2} that the shape of the line profile is likely at the origin of the correlated and anti-correlated lines. 

A "constant" asymmetric line profile cannot in itself produce an RV variation, the profile needs to vary with time. In the case of the lines that show an anti-correlation with stellar activity, the strong asymmetry is induced by blends. A symmetric change with time of these blends, or of the main line, will induce a spurious RV variation, which would not occur for symmetric line profiles, as demonstrated in Appendix~\ref{appendixA}. This has previously been shown by \citet{Reiners(2013)}, who studied the spurious RV effect in the near-IR (NIR) induced by Zeeman broadening, which is known to increase the width of spectral lines that are sensitive to a magnetic field without any asymmetry. We note that Zeeman broadening is extremely weak in the visible for magnetically quiet stars like the Sun \citep[e.g.][]{Robinson:1980aa, Anderson:2010aa}, even for lines with a large Land\'e factor, because the effect is proportional to the square of the wavelength. Therefore, Zeeman broadening is likely not the cause of the spurious RV observed in the anti-correlated lines. However, a spectral line could change in width and depth with activity as a result of a high sensitivity to temperature, which seems to have been observed in \citet{Davis(2017)} and \citet{Wise(2018)}.
 \citet{Brandt-1990} also observed a change in line width and depth due to stellar activity in resolved solar observations and discussed different physical processes that might explain this change in line profile morphology. We are currently investigating whether the anti-correlated spectral lines are more affected by a change in width or in depth, which could help to understand the origin of the effect. This work will appear in a forthcoming paper (Cretignier et al in prep).

\section{Detecting the inhibition of convective blueshift induced by stellar activity on single spectral lines}
\label{CB_inhibition}

As discussed in the introduction (see Sect.~\ref{introduction}), the goal of this paper is to demonstrate that the effect of the inhibition of convective blueshift in active regions can be detected in the $RV_{i}$ of individual spectral lines. If this is the case, we should observe that the sensitivity of the spectral line to stellar activity depends inversely on the line depth. Before we examine this, we first have to clean our selection of spectral lines to prevent spurious RV signals. We demonstrate
in Appendix~\ref{appendixA} and Sect.~\ref{RV_per_line_group} that a symmetric variation of an asymmetric line profile induces a spurious RV variation. Therefore, as discussed in the following section, we first rejected from our analysis all lines presenting a significant asymmetry.

\subsection{Detecting asymmetric lines} \label{remove_blends}

We followed two approaches to flag asymmetric line profiles. The first method consists of measuring seven different morphological parameters, derived from the line profile, its first and second derivative, in each spectral line, and define some cutoffs to reject the asymmetric line profiles. The morphological parameters we used with their respective cutoffs are detailed in Appendix~\ref{appendixAA}. The second method consists of verifying that no blend in the VALD 3 line list database \citep{Piskunov(1995),Kupka(2000),Ryabchikova(2015)} contaminated the selected line profile for each spectral line. Line depths were obtained from VALD3 by selecting the closest stellar parameters to $\alpha$ Cen B ($T_{\mathrm{eff}}=5250\,K$, $\log\,g=4.5$, macro-turbulence = 2\,\kms). The threshold for contamination was set at a depth ratio of 1\%. The first method allows us to split spectral lines into a symmetric and asymmetric group, and the second method selected an unblended and blended group.

Out of the 5753 spectral lines available, 2679 are found to be unblended and 1893 are classified as symmetric. A total of 1015 spectral lines are both at the same time.


\subsection{Detection of lines affected by convective blueshift inhibition}

As discussed in the introduction (see Sect.~\ref{introduction}), lines affected by the inhibition of convective blueshift should be strongly correlated with activity proxies. We therefore selected only lines
for which the Pearson correlation coefficient $\mathcal{R}$ between $RV_i$ and $RV_m$ was higher than 0.3 at $2\sigma$. 
As discussed in Sect.~\ref{line_by_line_RV}, we preferred
to study the correlation between $RV_i$ and $RV_m$ rather than $RV_i$ and \logrhk\, because this latter activity proxy probes the stellar chromosphere, and spectral lines are formed in the stellar atmosphere.

\begin{figure*}[h]
        \centering
        \includegraphics[width=18.7cm]{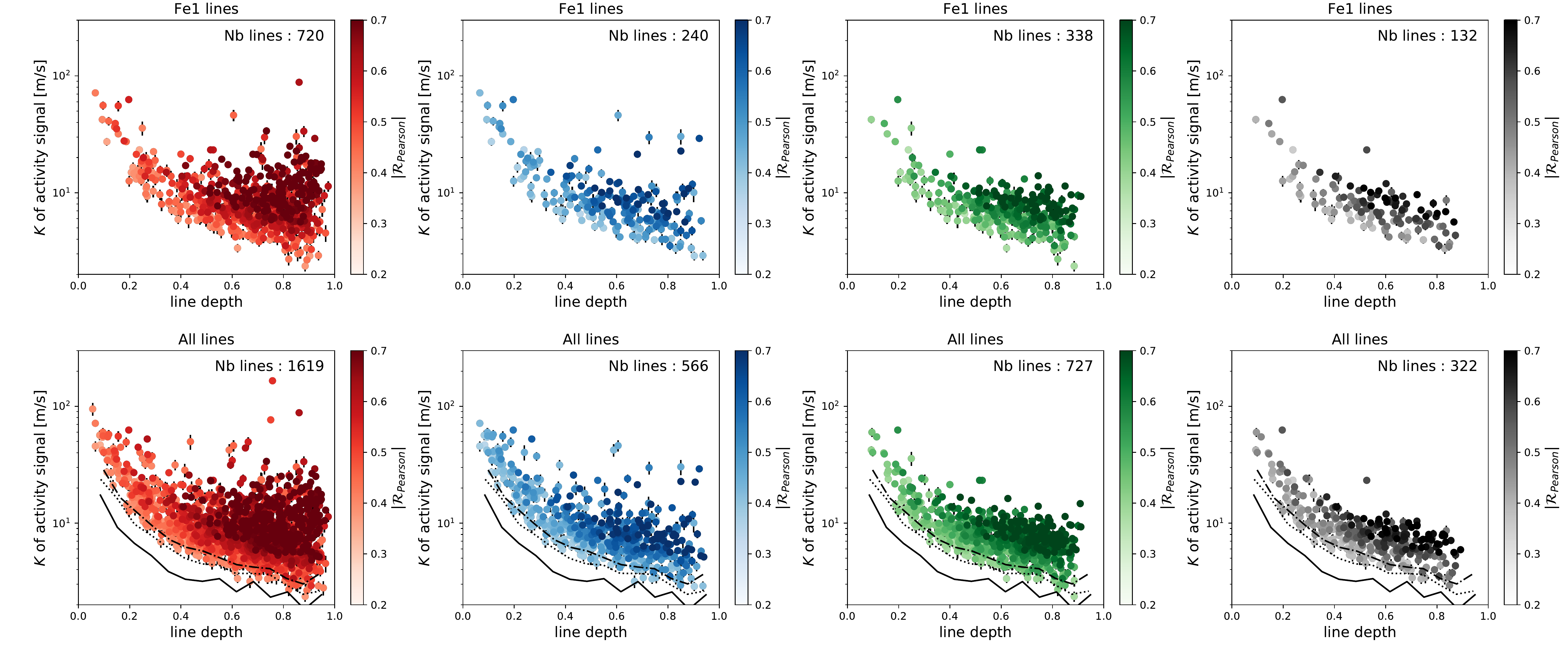}
        \caption{Radial velocity semi-amplitude $K$ induced by stellar activity ($K$ \textit{of activity signal} defined as the slope of the $RV_i$ vs $RV_m$ correlation times the RV semi-amplitude observed in the $RV_m$) as a function of the line depth for our four selection of lines: correlated lines (red dots),  unblended (blue dots), symmetric (green dots), and symmetric-unblended (black dots). The colour encodes the Pearson coefficient $\mathcal{R}$. The different curves, from bottom to top, represent the lower envelope of the correlated lines (left bottom panel) taking as threshold $\mathcal{R}=0.2, 0.3,$ and $0.4$, respectively. \textbf{Top}: 
                Fe I lines. \textbf{Bottom}: All species.}
        \label{FigReiners1}
\end{figure*}

In Fig.~\ref{FigReiners1} we display the slope of the correlation between $RV_i$ and $RV_m$ times the semi-amplitude of the activity signal seen in $RV_m$ (5\ms) as a function of line depth. Therefore, this is at first order equivalent to the semi-amplitude of the activity signal seen in $RV_i$ as a function of line depth. We show the results when only the Fe I spectral lines (top panels) or all the lines (bottom panels) are analysed. From left to right, we show the change in semi-amplitude of the activity signal seen in $RV_i$  as a function of line depth when we consider: i) correlated lines, ii) correlated unblended lines, iii) correlated symmetric lines and iv) correlated symmetric-unblended lines, with unblended and symmetric lines as defined in the preceding Sect.~\ref{remove_blends}. In all correlated lines, a significant number of strong spectral lines exhibit a large semi-amplitude for the activity signal, which implies that the $RV_i$ of these spectral lines seems strongly affected by stellar activity. However, in the results that only consider unblended, symmetric, and symmetric-unblended lines, the strong lines that seem to be significantly affected by stellar activity disappear. This implies that the RV variation of these lines is contaminated by a spurious RV effect induced by line profile asymmetry.

\begin{figure*}[h]
        \centering
        \includegraphics[width=18.7cm]{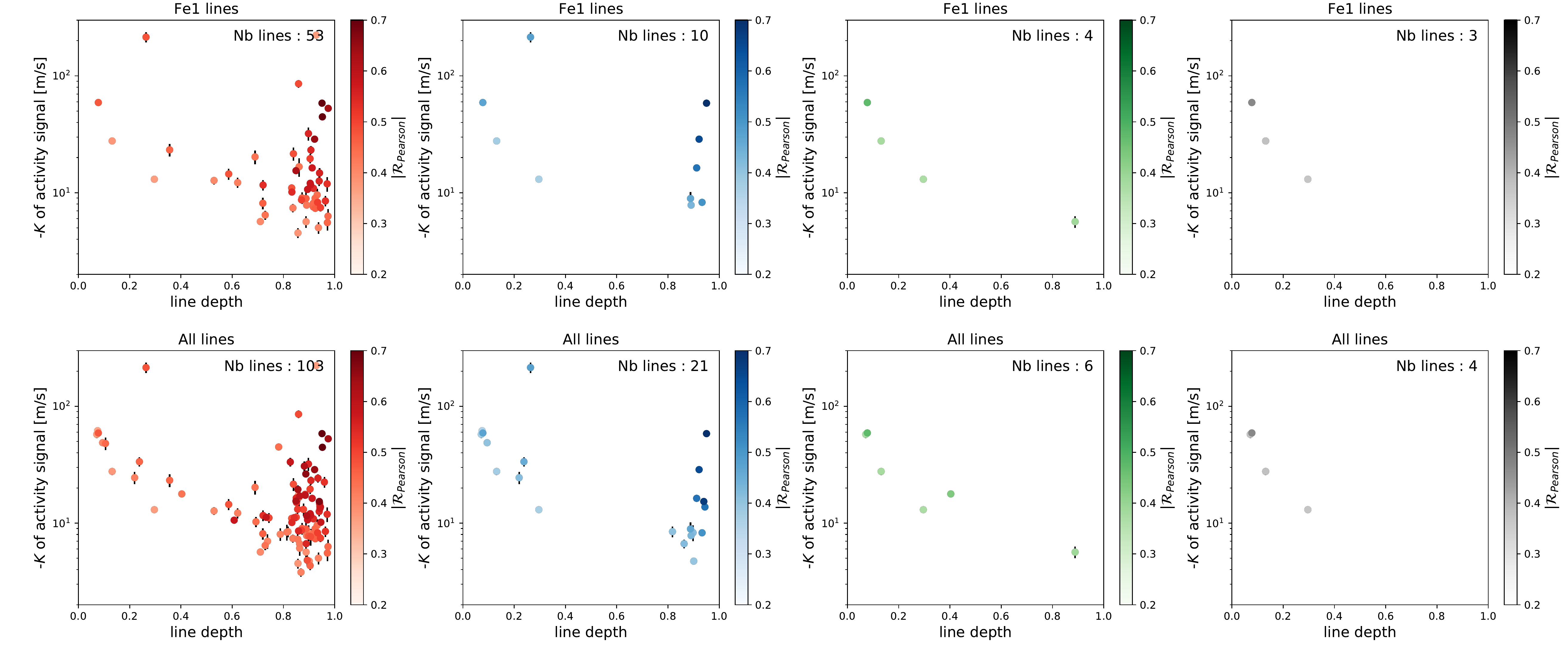}
        \caption{Same as Fig.~\ref{FigReiners1}, but with the anti-correlated lines ($\mathcal{R}<-0.3$ at $2\sigma$). Most anti-correlated lines are strong, asymmetric, and blended at the same time.}
        \label{FigReiners2}
\end{figure*}

To confirm that our method for rejecting asymmetric line profiles defined in Sect.~\ref{remove_blends} and Appendix~\ref{appendixAA} is efficient, we show in Fig.~\ref{FigReiners2} the same analysis as performed in Fig.~\ref{FigReiners1}, but with lines that exhibit a significant anti-correlation ($\mathcal{R}<-0.3$ at $2\sigma$). With the exception of a few lines, all disappear when only unblended, symmetric, or symmetric-unblended lines are selected. Only four symmetric-unblended shallow lines exhibit a significant anti-correlation out of the initial sample of 103 lines. We are therefore confident that our selection of symmetric-unblended lines removes the majority of spectral lines that are affected by a spurious RV effect that is induced by line profile asymmetry.

The results for the symmetric-unblended lines in Fig.~\ref{FigReiners1}, for which we are confident that we measured a true RV variation correlated with stellar activity, clearly show that shallow spectral lines present an activity signal whose semi-amplitude is much larger than that of the strong spectral lines. This is expected because shallow spectral lines are strongly affected by convective blueshift, which is not the case for strong spectral lines \citep[see e.g. Fig.~3 in ][]{Reiners(2016)}. Therefore, the inhibition of the convective blueshift observed in magnetic regions \citep[e.g.][]{Meunier(2010),Hanslmeier:1991aa,Schmidt:1988aa,Title:1987aa,Cavallini(1985)} can strongly affect the $RV_i$ of shallow spectral lines, which is not the case for strong spectral lines. Our results are in line with this picture. When the active region on $\alpha$ Cen B that is responsible for the observed RV variation is not visible on the stellar surface, shallow spectral lines are strongly blueshifted as a result of convection, which is no longer the case when the active region is visible and its internal magnetic field strongly inhibits convection. This leads to a strong variability in the $RV_i$ of shallow spectral lines. For strong spectral lines, the convective blueshift is much weaker, leading to a lower variability. This behaviour has been simulated by \citet{Meunier:2017aa}, but this is the first time that we are able to measure it in real observations.

Although the observed dependency of the activity signal amplitude as a function of line depth in Fig.~\ref{FigReiners1} for the unblended-symmetric selection can simply be explained by the difference in velocity of the convection as a function of depth inside the stellar atmosphere and the inhibition of convective blueshift in active regions, biases might be a concern when the spectral lines shown in Fig.~\ref{FigReiners1} are selected. In each panel of this figure, the colour coding corresponds to the Pearson correlation coefficient $\mathcal{R}$ between $RV_i$ and $RV_m$. We also show the lower envelope of the points when lines with a correlation coefficient $\mathcal{R}$ larger than 0.2, 0.3, and 0.4 at $2\sigma$ are considered. With this additional information, we clearly see that the lower envelope of the points can be explained by the bias due to noise when the correlation coefficient $\mathcal{R}$ was estimated. Shallow spectral lines have a lower RV precision, therefore these lines need to present a strong RV signal induced by stellar activity to satisfy the cutoff imposed in $\mathcal{R}$. However, the upper envelope of the points also shows the dependency of the activity signal amplitude on line depth, and we were not able to explain this envelope by any bias.

In the upper and lower panel of Fig.~\ref{FigReiners1}, we show the results for  the Fe I spectral lines alone and for all the lines, respectively.
At first order, we expect to observe a similar behaviour for the Fe I compared to all the other lines because all lines are produced in a relatively small 
region of the stellar atmosphere. We expect second-order differences, however, but they seem to be buried in the noise.

\subsection{Radial velocity semi-amplitude induced by the convective blueshift inhibition as a function of line depth}

\begin{figure*}[h]
        \centering
        \includegraphics[width=18cm]{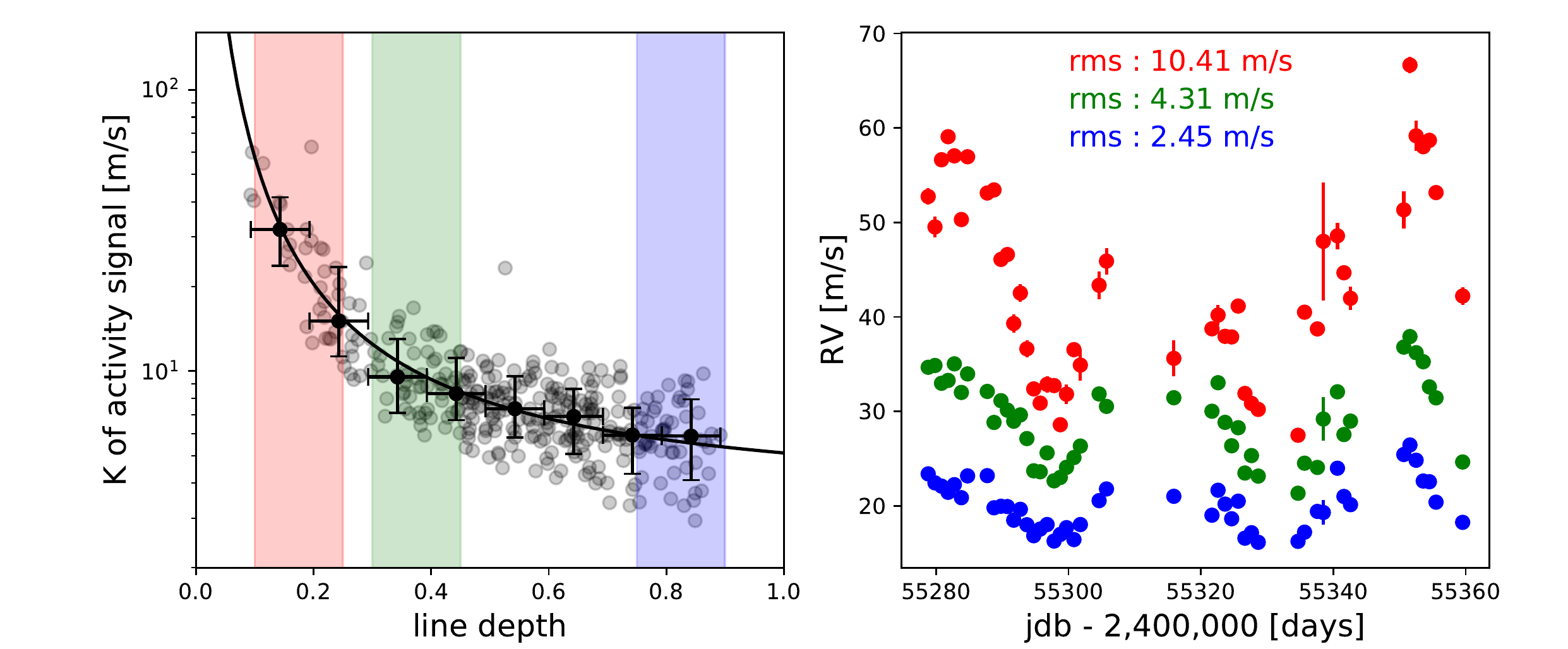}
        \caption{\textbf{Left}: Radial velocity semi-amplitude of the stellar activity signal as a function of the line depth $d$ for the correlated symmetric-unblended lines (grey dots). The data are binned into nine bins of size 0.1 (black dots), and the best fit for a second-order polynomial in $1/d$ is shown by the black line. \textbf{Right}: Radial velocity of the subselections of lines falling in the area of the left plot panel. The RV amplitude grows from strong (blue) to shallow (red) lines.}
        \label{FigReiners3}
\end{figure*}
In this section, we focus on the correlated symmetric-unblended lines, for which we are confident that the RV effect is mainly induced by the convective blueshift inhibition and thus by stellar activity. Fig.~\ref{FigReiners1} clearly shows that the RV semi-amplitude is inversely proportional to the line depth $d$. In Fig.~\ref{FigReiners3} we fit the observed relation with a second-order polynomial in 1/$d$, which was the degree that minimised the Bayesian information criterion. The best fit is given by the following equation:
\begin{equation}\label{eq:rv_cb}
RV_{CB} \: [m/s] =  3.3 + \frac{1.4}{d}+ \frac{0.40}{d^2}.
\end{equation}
This relation is only valid for $\alpha$ Cen B. For other stars, according to the results shown in \citet{Gray(2009)},
we expect that the velocity of the convective blueshift as a function of line depth is similar and only depends on a scaling factor. \citet{Meunier(2017)} showed 
from simulations that a multi-linear relation between the convective blueshift velocity, the line depth, the effective stellar temperature, and the chromospheric activity index is expected. They also reported that the attenuation factor of the convective blueshift velocity with activity seems independent of the stellar type for G and K dwarfs.

\begin{figure}[tp]
        \centering
        \includegraphics[width=9.3cm]{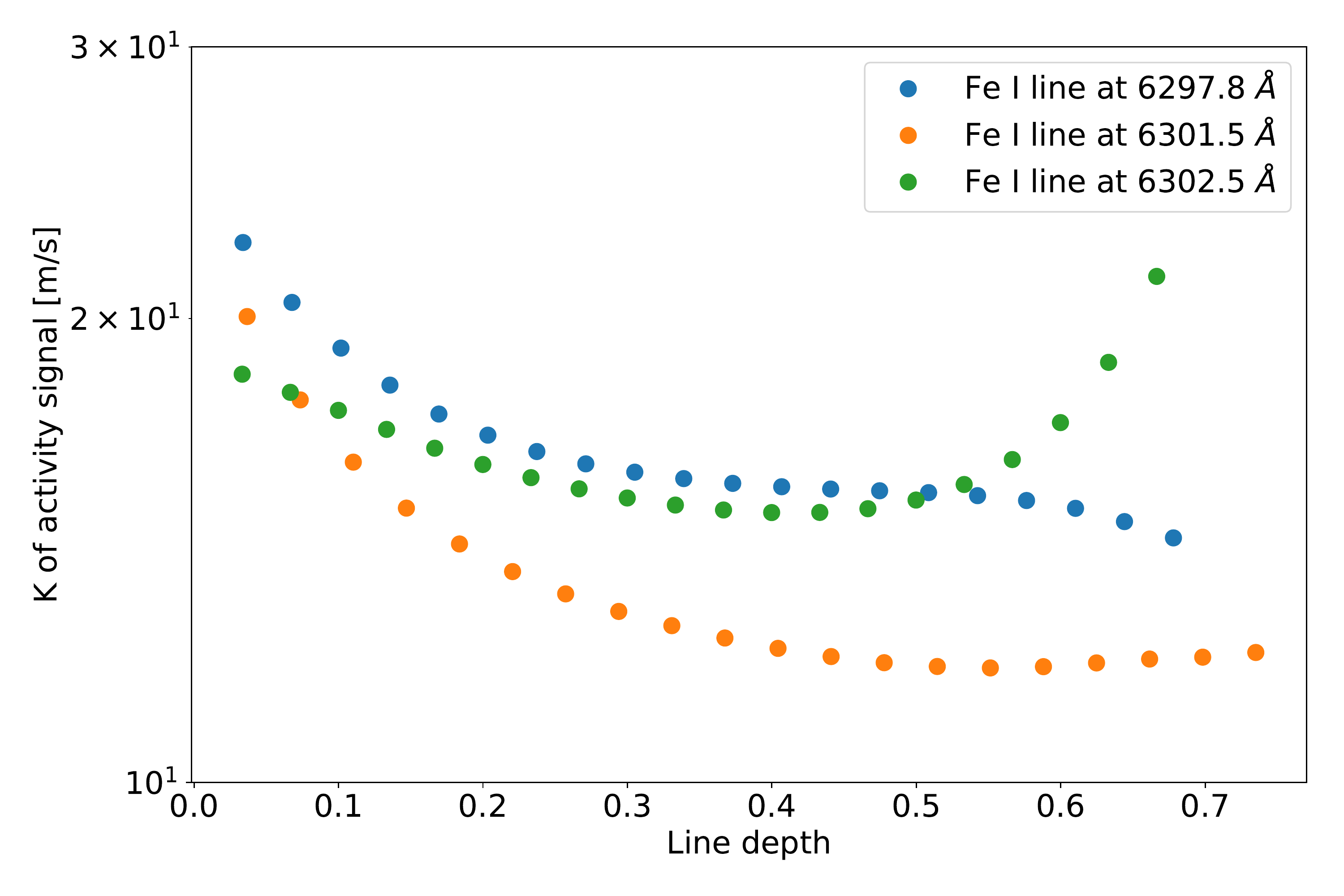}
        \caption{Radial velocity semi-amplitude of the activity signal vs. line depth simulated with SOAP\,2.0 and for the three solar Fe I 
                spectral lines at 6297.8, 6301.5, and 6302.5\,$\AA$ analysed in \citet{Cavallini(1985)}. This figure can be compared to Fig.~\ref{FigReiners3}. The behaviour as a function of line depth can be attributed to the inhibition of the convective blueshift in solar facula regions.}
        \label{FigSimu}
\end{figure}

As a first guess, we would expect to observe in Fig.~\ref{FigReiners3} a similar behaviour of the convective blueshift velocity as a function of line depth relation shown in Fig.~3 in \citet{Reiners(2016)}. However, the authors there estimated the convective blueshift velocity by measuring the shift between the line core wavelength and the wavelength of the line measured in the laboratory. In our case, we measured how the convective blueshift is inhibited by stellar activity. In addition, we measured the RV on the entire line profile and not only on the line core. Therefore, to show that the observed trend in Fig.~\ref{FigReiners3} is physical, we used the SOAP\,2.0 code \citep[][]{Dumusque(2014)} to simulate the expected behaviour for the three spectral lines measured at extremely high resolution by \citet{Cavallini(1985)} in a solar facula. This simulation is described in detail in Appendix~\ref{appendixC}, and the results are shown in Fig.~\ref{FigSimu}. The result of this simulation based on solar observations is consistent with what is observed for $\alpha$ Cen B in Fig.~\ref{FigReiners3}.

\subsection{Mitigating the stellar activity signal}

We demonstrated that shallow spectral lines present a stronger RV signal induced by stellar activity than strong spectral lines, therefore we expect that removing the shallow lines when the RV of a star is computed mitigates the effect from activity. Unfortunately, the precision of the RV measured on shallow spectral lines is lower than for strong spectral lines, therefore the RV of shallow lines does not contribute significantly to the
$RV_m$ measured on all the lines. We find an RV rms of 3.12\ms when we calculate the $RV_m$ of $\alpha$ Cen B using only the symmetric-unblended group of lines. Splitting this group into two with lines shallower and stronger than 0.5, we obtain an RV rms of 4.36 and 2.86\ms, respectively. This shows that the RV rms obtained with the strong lines is only slightly lower, by 0.26\ms\, , than the RV rms when all the symmetric-unblended lines are considered. However, there is a significant difference between the two RV rms, with the $RV_m$ measured on shallow spectral lines exhibiting an activity signal 1.5 times stronger. \citet{Meunier(2010)} compared only the velocity of the convective blueshift between shallow and strong spectral lines and expected an activity signal seen in RV 2.7 times higher for shallow spectral lines. This estimate is for the Sun, and we consider that the convective velocity for a K1 dwarf such as $\alpha$ Cen B should be twice as low \citep{Beeck(2013a)}, consistent with the ratio of 1.5 that we obtain here.

A significance difference is observed between the RVs measured on shallow and strong spectral lines,
therefore the difference between these two RV data sets should give us an interesting stellar activity proxy. In addition, because a planetary signal will affect the RV measured on different subsets of spectral lines in the same way,
this new activity proxy is free of any planetary signals. We can therefore use it to model and mitigate the activity signal in the $RV_i$ of each individual spectral line by fitting a scaled version of this new stellar activity proxy, 
and then performing a weighted average
over all the $RV_i$ to obtain the $RV_m$ corrected for the stellar activity signal. When this correction is performed and the subset of correlated symmetric-unblended spectral lines consists of those that are shallower and stronger than 0.50, we obtain an rms for the corrected $RV_m$ of 1.42 \ms, 
2.2 times smaller than the rms of 3.12 \ms \ that we computed when we averaged the RV information of all the correlated symmetric-unblended spectral lines. In theory, a library of activity proxies can be produced by taking different thresholds in depth $d$, where the activity proxy can be defined as the RV of the group shallower than $d$ minus the RV of the group stronger than $d$. All these proxies could then be linearly decorrelated to mitigate the stellar activity signal. For instance, fitting two components on the $RV_m$, using as thresholds in depth $d=0.50$ and $d=0.38$ based on the correlated line selection, leads to a rms of 0.84 \ms. 
We will investigate a more sophisticated analysis of stellar activity mitigation in a forthcoming paper.


\section{Conclusions}\label{conclu}

Following  \citet{Dumusque(2018)}, we studied in detail the RV of each individual line in the visible spectra of $\alpha$ Cen B in 2010, with the purpose
of measuring for each spectral line the effect of convective blueshift inhibition induced by stellar activity.
To better measure the RV of the individual spectral line, we first optimised the windows and line centres directly on a master spectrum of $\alpha$ Cen B.  \citet{Dumusque(2018)} selected line centres based on the values from the HARPS K5 mask used to perform cross-correlation for K dwarfs, and the window width was a fixed number of pixels. It is clear that the choice made in \citet{Dumusque(2018)} 
was not optimal because the windows of certain spectral lines overlapped, which created some redundancy in the estimate of the RV, in addition to being sensitive to the RV of 
neighbouring spectral lines. We also optimised the line selection to be less affected by telluric line contamination. We cross-matched all the lines found with a model for the Earth atmosphere absorption, and rejected lines that
were contaminated by more than 2\%. Compared to the selection of lines performed in \citet{Dumusque(2018)}, our final selection of spectral lines contains 1.45 time more lines for a total RV information that is 9\% greater.

We then investigated the effect of the shape of the line profile on the measured RV. We demonstrate in Appendix
\ref{appendixA} that a symmetric variation, like a change in depth or width, of an asymmetric line profile induces a spurious RV variation, which has previously been speculated in \citet{Reiners(2013)}. 
Because it was observed in the Sun that spectral lines change in depth and width due to stellar activity \citep[e.g.][]{Brandt-1990}, measuring the RV of spectral lines presenting 
an asymmetric line profile can lead to a spurious RV effect that is correlated with stellar activity.

By selecting spectral lines with a symmetric line profile, for which we are confident that the derived RV is only affected by a Doppler effect and not by a spurious signal induced by line profile  asymmetry, we were able to show for the first time that
the amplitude of the RV signal induced by stellar activity is inversely proportional to the line depth, where the fitted coefficients are found in equation (\ref{eq:rv_cb}). This can be explained from a physical point of view by the fact that shallow spectral lines, formed deep inside the stellar photosphere, are strongly affected by the inhibition of the convective blueshift that is observed inside active regions. Strong spectral lines, whose core is formed close to the stellar surface, are much less affected. These results are in agreement with the fact that the velocity of the convective blueshift is stronger deep inside the photosphere than close to the stellar surface \citep[e.g.][]{Reiners(2016), Dravins-1981}. In addition, we were also able to show in Appendix \ref{appendixA} that for spectral lines for which we obtain a good enough precision and for which we are sure that the measured RV is induced by a Doppler shift (symmetric spectral lines stronger than 0.5), at least 89\% of them are affected by the inhibition of the convective blueshift.

Finally, we investigated a few simple possibilities to mitigate stellar activity based on the relation between strength of the activity effect and line depth. By splitting the correlated lines into two groups, those shallower and stronger than a threshold $d$, we were able to mitigate stellar activity from a level of 2.17\ms\,down to 0.84 \ms.

In further works, we will investigate whether the same conclusions holds for solar observations obtained with the HARPS-N solar telescope \citep[][]{Collier-Cameron:2019aa, Dumusque(2015)} and will present more complex techniques to mitigate stellar signals in the presence of planetary signal. The most promising method probably is the formation of a new activity proxy formed by different line selections, which could theoretically be done for any star, without being limited by photon noise.  

\begin{acknowledgements}
We thank the Swiss National Science Foundation (SNSF) and the Geneva University for their continuous support. We want to acknowledge the fruitful, in-depth discussions that took place at ISSI (International Space Science Institute) in Bern during the meeting "Towards Earth-like alien worlds: Know thy star, know thy planet" with an ISSI International Team, December 10–14, 2018. We are also grateful for the help of Nathan Hara and its suggestion to compute uncertainties on the Pearson coefficients. This work has been in particular carried out in the framework of the \emph{PlanetS} National Centre for Competence in Research (NCCR) supported by the SNSF. X.D is grateful to the Branco-Weiss Fellowship--Society in Science for its continuous financial support.

\end{acknowledgements}

%
%

\bibliographystyle{aa}
\bibliography{Paper_Cretignier_2019_corrected}

\begin{appendix} 

\section{New window selection for spectral lines}
\label{appendixAA}

To mitigate the RV effect relative to blends and inadequate window selection for spectral lines, we performed a tailored selection of lines and windows based on the observed spectra of $\alpha$ Cen B. The first step consists of building a high-resolution spectrum of the star by stacking individual spectra. To prevent perturbation by stellar
activity that modifies the shape of spectral lines, out of the 1767 1D HARPS spectra available for $\alpha$ Cen B in 2010, we selected the 10 percentile spectra at the lowest level of stellar activity.
The 1D HARPS spectra were first corrected for the binary motion of $\alpha$ Cen B around $\alpha$ Cen A, which amounts to 40\ms in
2010 (see the middle panel of Fig.~\ref{FigRvRHK}), before we  interpolated the spectra on a common wavelength grid between 3781.77 and 6912.68\,$\AA$ with a step of 2 $m\AA$ using a cubic spline.
We then stacked all the spectra, which provided a master quiet spectrum (MQS), and shifted the entire spectrum by 22.7 km/s to remove the systemic velocity of the $\alpha$ Cen stellar system \citep{Wesselink(1953)}. The continuum of the MQS was estimated by a rolling maximum with a window size of 10 $\AA$. The sub-product of this operation produces a step-like function. In order to smooth this function, only one point was kept every 2 $\AA,$ and points that were visually too low because of broad absorption lines in the original MQS were removed. Finally, this continuum was interpolated with a cubic spline on the initial grid and used to normalise the spectrum.

To detect all the spectral lines, we first calculated all the local extrema of the MQS. The minima provide a sufficient estimate of the line centres (red points in Fig.~\ref{FigMask}),
whereas the two maxima on each side of a line give us the maximum window that could be used before probing a neighbouring line (green points in Fig.~\ref{FigMask}).
Selecting lines based on extrema yielded many spurious lines in the continuum due to noise because in this simple description, a line is just a minimum surrounded by two maxima. To remove spurious line,  we rejected all lines with a depth shallower than 0.05 and with a window width smaller than 10 pixels. 

After selecting the centre of the lines that we wished to study, we selected a spectral window surrounding each line on which the RV was calculated. The selection of this window
is critical to derive a reliable line-by-line RV because if the window is too large, the RV of neighbouring lines will be mixed together, and if the window is too small, the RV will only be computed on the core of
the line, which is poorer in Doppler information than the wings. For simplicity, we only chose a window that was symmetric with respect to its line centre. In this context, an
obvious window can be obtained by taking the minimum distance between the line centre and the two local maxima surrounding the line. This window, called large window (LW), is
the largest possible window without measuring information on adjacent lines. A small window (SW) can also be derived by selecting the negative local minima of the
second derivative of the MQS. These negative local minima lie farther out than the inflection points of the spectral lines, and therefore allow selecting a significant fraction of the line.
All the details are given in Appendix \ref{appendixB}. Finally, we define the medium window (MW) as the average between the SW and the LW.

\section{Spectral line RV anti-correlated with the stellar activity signal}
\label{appendixA}

\begin{figure*}[h]
        \centering
        \includegraphics[width=16.3cm]{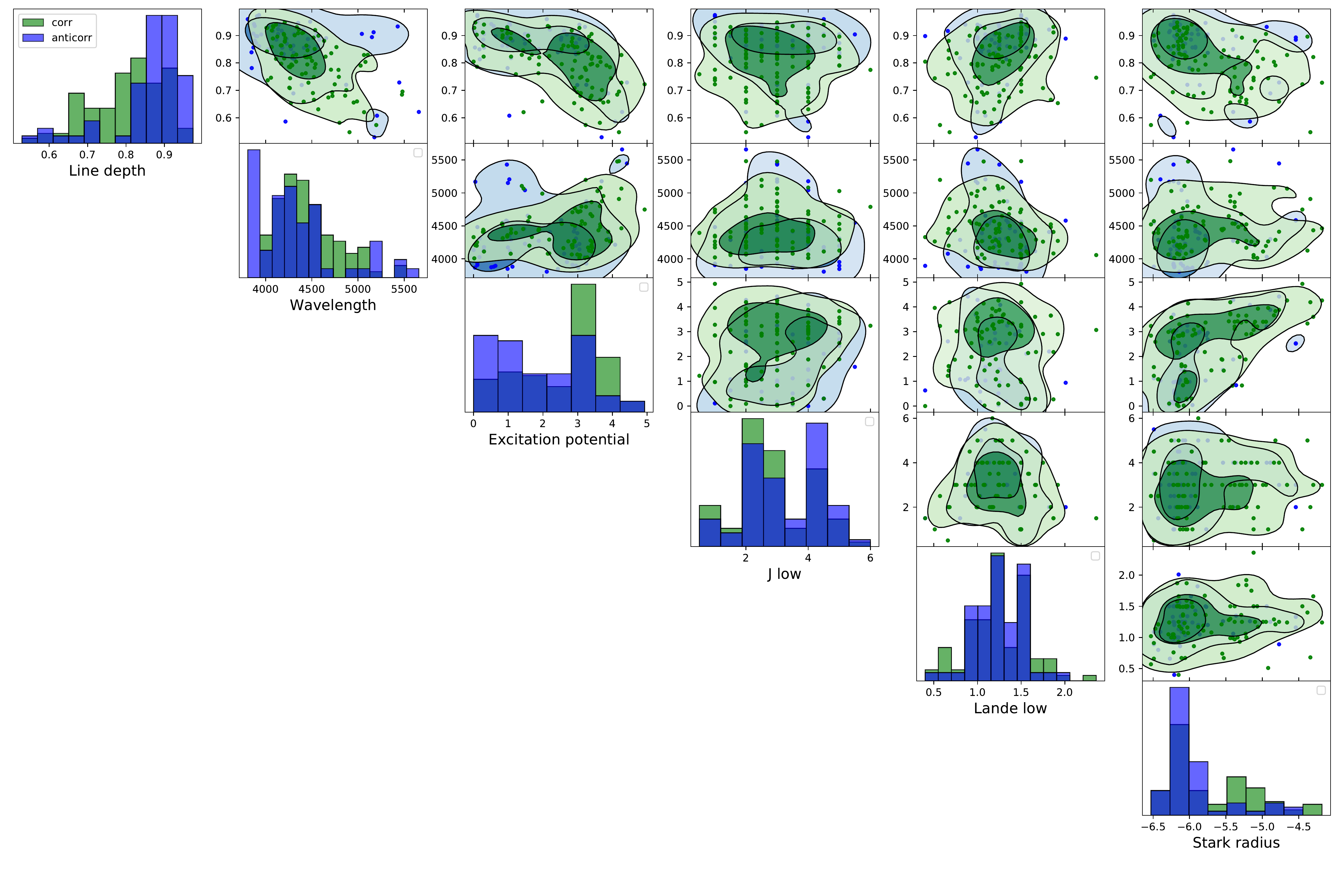}
        \caption{Corner plot of six physical line parameters (line depth, wavelength, excitation potential, angular quantum number J$_{\mathrm{low}}$ of the lowest energy level, Land\'e factor (Lande low), and Stark radius) for the correlated (\textit{green}) and anti-correlated (\textit{blue}) groups. Kernel density estimates at 1 and 2$\sigma$ are overplotted. No significant difference is visible between the two groups.}
        \label{FigKDE1}

        \centering
        \includegraphics[width=16.3cm]{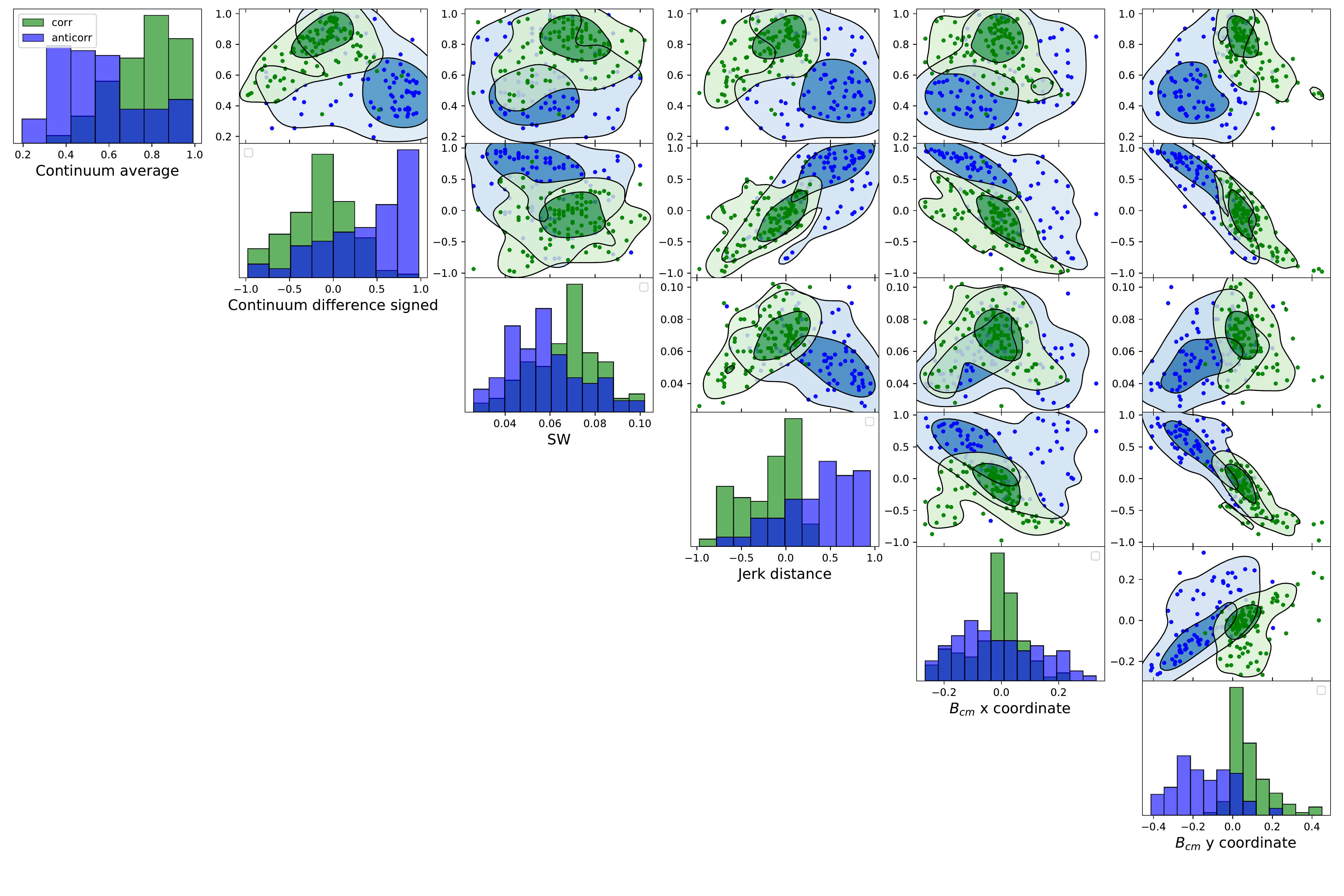}
        \caption{Same as Fig.\ref{FigKDE1} for six morphological parameters that probe the line profile asymmetry (continuum average, continuum difference with sign, small window (SW), Jerk distance, and barycentre of the derivative $B_{cm}$ in x and y coordinates). This time, the two groups are well separated in the parameter space.}
        \label{FigKDE2}
\end{figure*}

In this section, we investigate why the RV of certain spectral lines is anti-correlated with the stellar activity signal. Our first guess was that the anti-correlated lines are sensitive to a stellar activity signal originating from a different physical process. We therefore searched for differences in the physical parameters of these anti-correlated spectral lines. The two correlated and anti-correlated groups were balanced in size and were defined with a Pearson coefficient $\mathcal{R} > 0.7$ and $<-0.3$ at $2\sigma,$ respectively. To obtain the physical parameters of the spectral lines for which we had an RV measurement, we cross-matched our selection of lines with the closest VALD 3 \citep{Piskunov(1995),Kupka(2000),Ryabchikova(2015)} model for $\alpha$ Cen B ($T_{\mathrm{eff}}=5250\,K$, $\log\,g=4.5$, macro-turbulence = 2\,\kms). We then investigated whether the physical parameters of the lines could explain the correlated and anti-correlated behaviour. Fig.~\ref{FigKDE1} shows that none of the physical parameters can explain the correlated versus anti-correlated spectral lines. It seems therefore at first order that the physical properties of the lines are not the origin of the anti-correlated spectral lines.

We then tried to determine whether the morphology of the lines might explain the correlated and anti-correlated behaviour. In Appendix~\ref{appendixB} we define several morphological parameters that can be used to assess the asymmetry of a spectral line. Fig.~\ref{FigKDE2} shows that these morphological parameters can separate the correlated from the anti-correlated lines in parameter space. The anti-correlated lines therefore present a line profile that is significantly different from the correlated lines. All the anti-correlated spectral lines have an asymmetric line profile.

To understand why an asymmetric line profile induces an RV signal that is anti-correlated with stellar activity, we simulated the observed Rv considering three different effects:
a Doppler shift, a change in width that could be due to Zeeman broadening or a change in stellar atmospheric temperature gradient  \citep[e.g.][]{Reiners(2013), Brandt-1990}, and a depth modification as observed for temperature-sensitive lines
\citep[e.g.][]{Wise(2018), Thompson(2017), Gray(2008), Gray(1991)}. The two latter transformations should not introduce an RV shift a priori because they are symmetric with respect to the line centre.

\begin{figure*}[h]
        \centering
        \includegraphics[width=18.5cm]{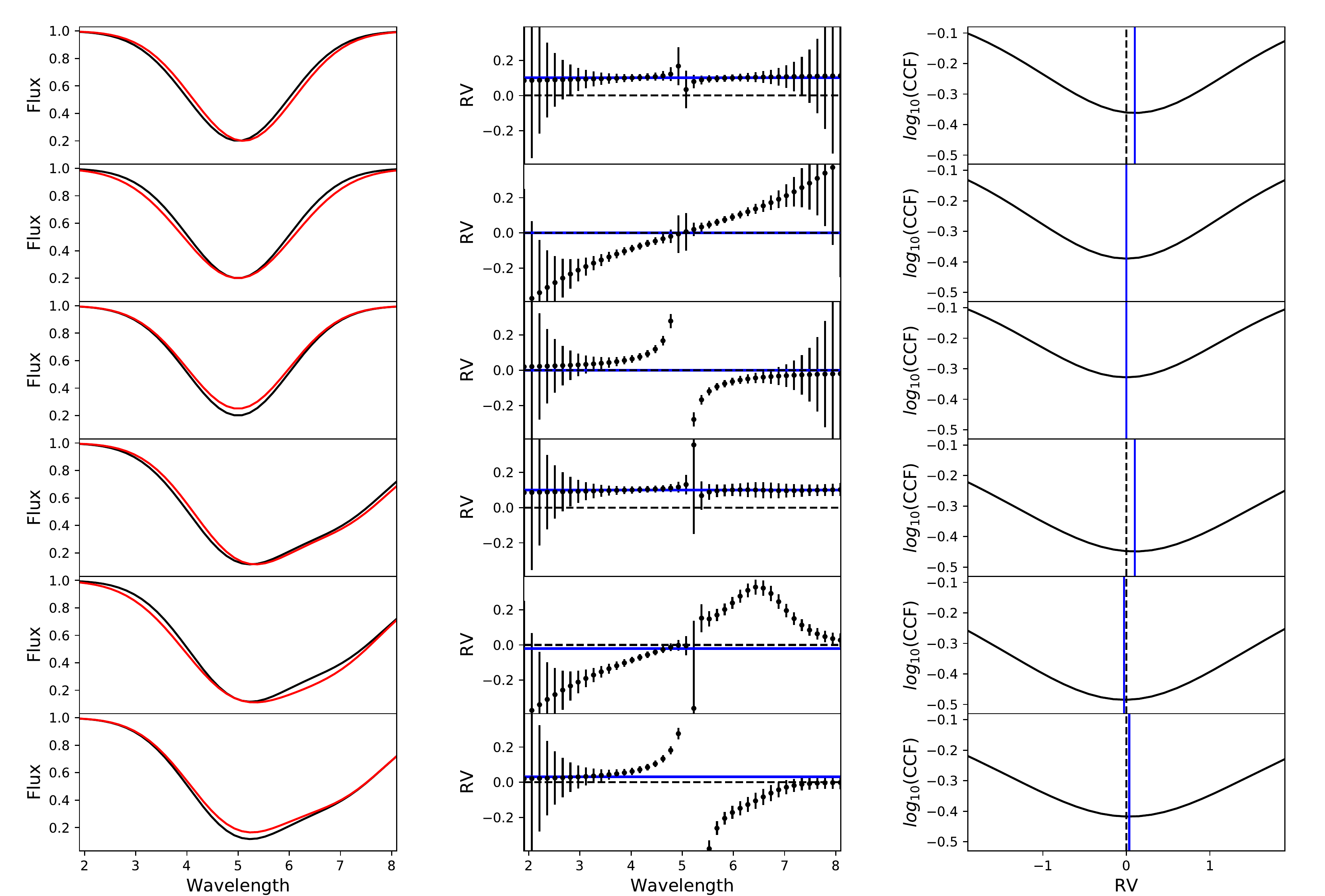}
        \caption{Simulation showing how symmetric flux variations can introduce a spurious RV shift for an asymmetric line profile. All the units are arbitrary. \textbf{Left column}: Initial line profile (black curve) compared to the profile modified by the active regions (red curve). \textbf{Middle column}: RV derivation according to the linearisation of the Doppler shift \citep{Bouchy(2001)}. The RV is the weighted average of the local RVs. The uncertainties are at first order inversely proportional to the derivative of the flux. The zero RV (\textit{dashed line}) is compared to the measured RV (blue line). \textbf{Right column}: RV derived using the barycentre of the CCF. \textbf{First row}: A 0.1 Doppler shift on a symmetric line profile, the Bouchy and CCF method extract the correct RV shift. \textbf{Second row}: Change in width of a line induced by Zeeman broadening, pressure broadening, or other effects \citep[][]{Brandt-1990}. The extracted RV shift is null because each modification of the left wing is compensated for by the same modification of the right wing. \textbf{Third row}: Depth modification induced by temperature-sensitive lines. Again, no RV shift is measured.  \textbf{Fourth, fifth, and sixth rows}:  Same as the three first rows, but a blend was inserted on the right side of the line. We note that in this case, the symmetric variations introduce a spurious RV shift. This RV shift would be inverted if the blend were inserted on the left wings instead of the right wing.}
        \label{FigSchema}
\end{figure*}

In the case of a symmetric line profile, Fig.~\ref{FigSchema} shows that a symmetric variation does not introduce a spurious RV shift. However, when the line profile is asymmetric, a spurious RV shift is observed. This effect is measured regardless of the method used to derive the RV, either the method used in this paper and presented in \citet{Bouchy(2001)}, or a cross-correlation and an RV that is defined as the centre of the Gaussian fit to the derived cross-correlation function.

The fact that a symmetric modification of an asymmetric line profile induces an RV variation is easy to understand.
Because the local RV over a spectral line profile is proportional to the derivative of the flux and its weight to the square of the flux derivative, see \citet{Bouchy(2001)}, any asymmetry induced by a blend will break the balance between
the left and right wing of a spectral line and therefore induce a spurious RV. 
With this simple model, it is easy to understand that the lines that present an RV that is anti-correlated with stellar activity are thus blended lines that present a depth variation that is correlated with stellar activity, likely temperature-sensitive lines, with a steeper right wing, or a blended line that presents a width variation that is correlated with stellar activity, likely due to Zeeman broadening or other effects, with a steeper left wing.

\begin{figure}[tp]
        \centering
        \includegraphics[width=9cm]{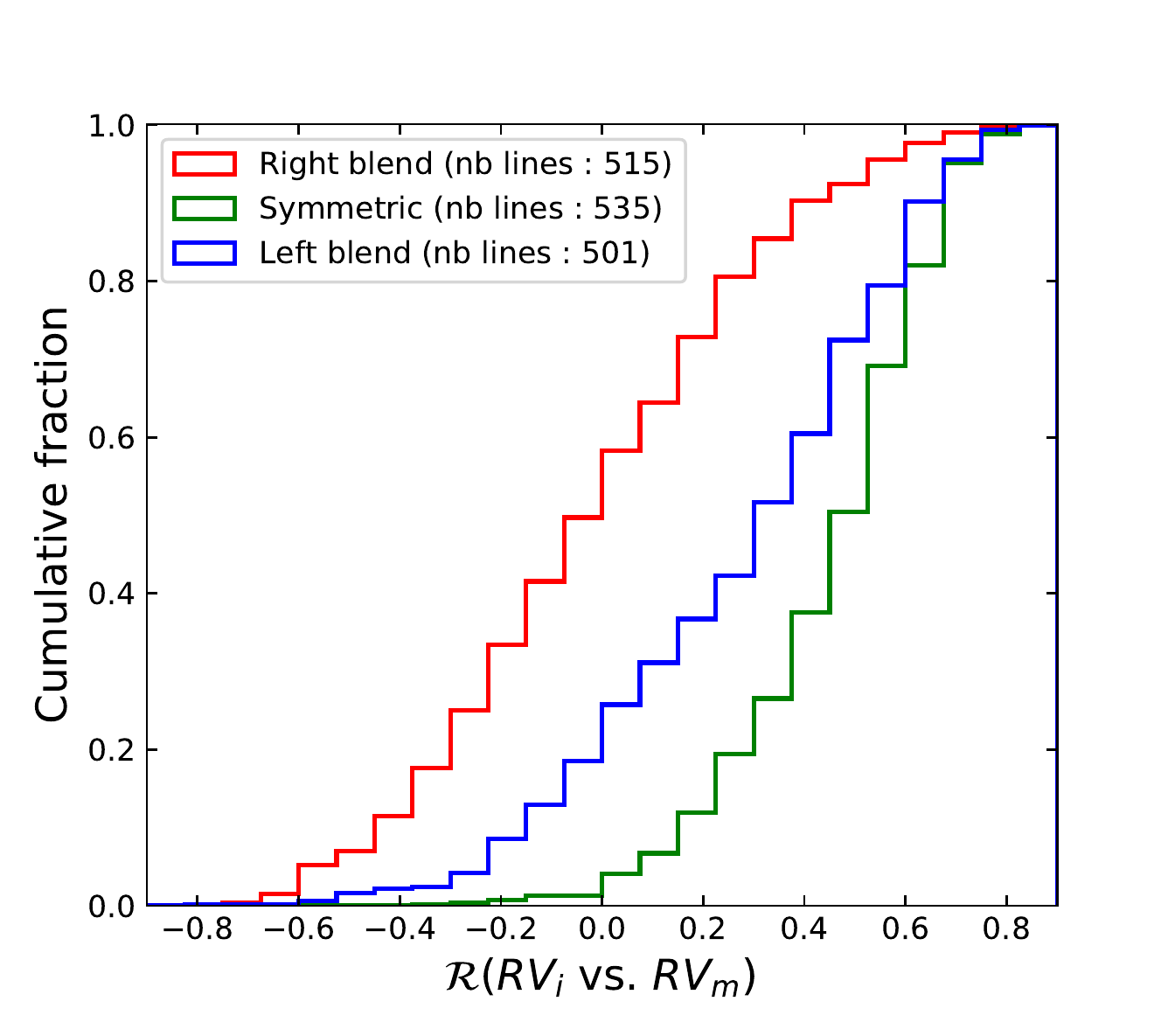}
        \caption{Cumulative distribution of the Pearson coefficient between the $RV_i$ and $RV_m$ correlation for three classes of lines: symmetric (green), right blended (red), and left blended (blue).}
        \label{Figblends3}
\end{figure}

In Fig.~\ref{Figblends3} we show the cumulative distribution of the coefficient of the correlation $\mathcal{R}$ for three classes of lines: symmetric, left-blended, and right-blended lines, where a left- or right-blended line lacks a left or right wing, respectively. We took the same criterion for symmetry as presented in Appendix~\ref{appendixB}, except that we were slightly more restrictive and used an absolute continuum difference smaller than 0.15. Left- and right-blended lines were defined with the criterion of a continuum difference larger than 0.4 and smaller than $-0.4$, respectively. We finally only considered spectral lines with a depth larger than 0.5 for which the RV precision is sufficient to provide a trustworthy correlation coefficient.
 We observe that most of the anti-correlated lines are right blended and nearly none are in the symmetric class. Also, most of the symmetric lines possess a low correlation with stellar activity. We counted that $\sim89\%$ of these symmetric lines possess a coefficient   $\mathcal{R}>0.30$ at $2\sigma$.  Because an RV signal measured from these lines can only be due to a Doppler shift, this indicates that most of the lines in the stellar spectrum are affected by the convective blueshift inhibition induced by stellar activity.

\begin{figure*}[h]
        \centering
        \includegraphics[width=18cm]{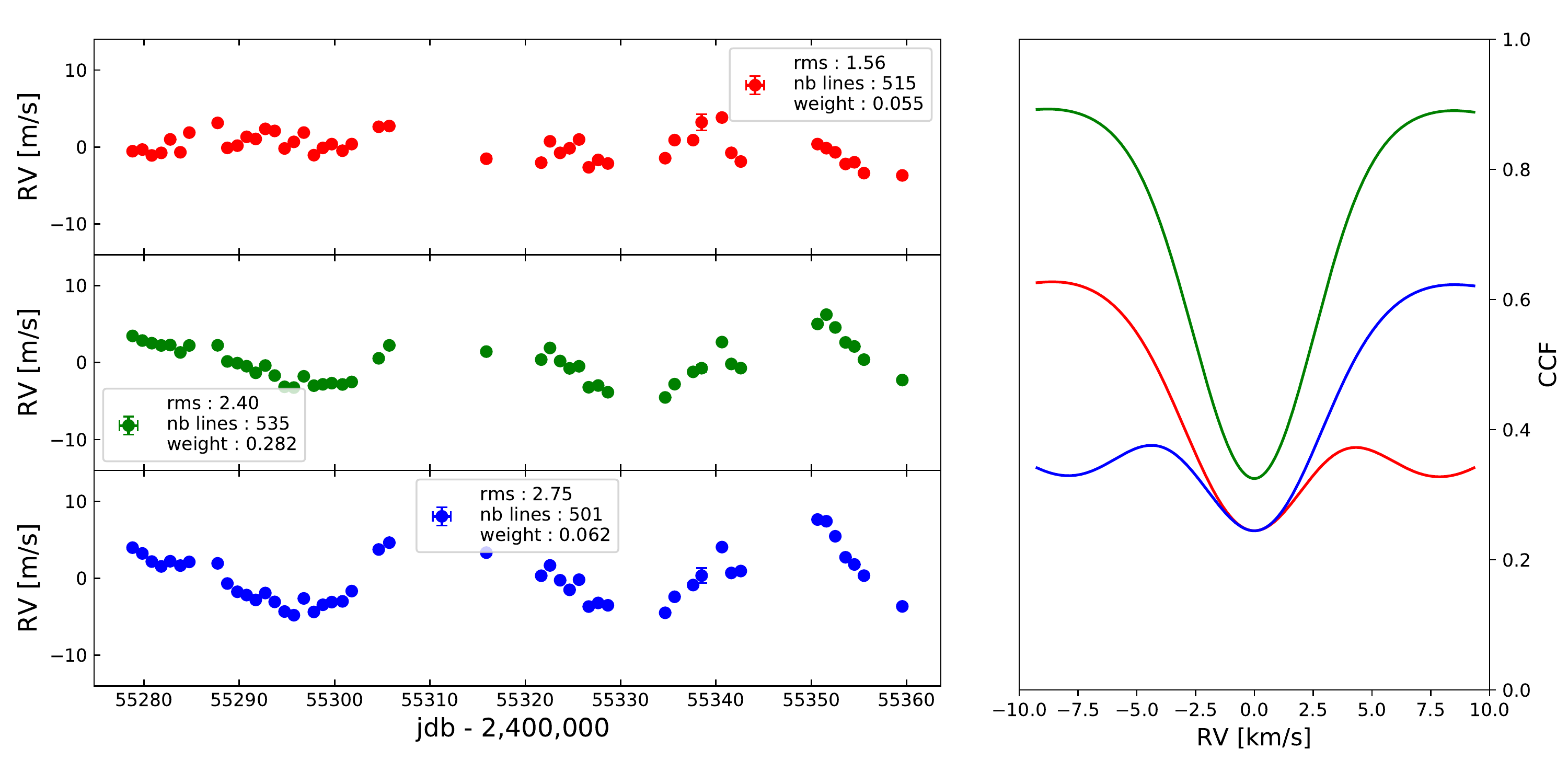}
        \caption{\textbf{Left}: $RV_m$ measured in the different subgroups of left-blended, right-blended, and symmetric lines shown in Fig.~\ref{Figblends3}. \textbf{Right}: CCF for the three different line groups obtained after cross-correlating the spectrum with a binary unweighted mask centred on line positions.}
        \label{Figblends3_rv}
\end{figure*}

In Fig.~\ref{Figblends3_rv} we show in the left panel the $RV_m$ calculated for each subgroup of right-blended, left-blended, and symmetric lines and in the right panel, the CCF for each of these subgroups, estimated using an unweighted binary mask with ones at the position of line centres and zeros elsewhere. The asymmetry observed in the CCF for the right- or left-blended lines is striking. The $RV_m$ measured on the
right-blended lines present an rms that is smaller than the symmetric lines, and the opposite holds for the left-blended lines. The reason is that for right-blended lines, the anti-correlated spurious RV effect with stellar activity induced by
line profile asymmetry compensates for the convective blueshift inhibition effect, which is correlated with stellar activity. For left-blended lines, the correlated spurious RV effect with stellar activity induced by this specific line profile asymmetry sums with the convective blueshift inhibition effect, which finally yields larger RV rms.

In conclusion, these are two different phenomena that induce an RV effect. On the one hand, modification of the depth or width of the spectral lines induced by stellar activity, which creates a spurious RV effect when spectral lines present an asymmetric profile. Depending on the shape of the asymmetry, this spurious RV effect can be correlated or anti-correlated with stellar activity. On the other hand, convective blueshift inhibition that produces a redshift of spectral lines, and therefore RVs that are correlated with stellar activity. It is hard to decorrelate one effect from the other, but by selecting only symmetric line profiles, we should be sensitive only to convective blueshift inhibition.

\section{Morphological parameters for diagnosing line profile asymmetry}
\label{appendixB}
In this section, we define some parameters that can be used to study the morphology of a spectral line. They were used either to select the windows in which we measured the RV of
a spectral line or to distinguish between symmetric or asymmetric spectral lines.

\begin{figure}[h]
        \centering
        \includegraphics[width=9.3cm]{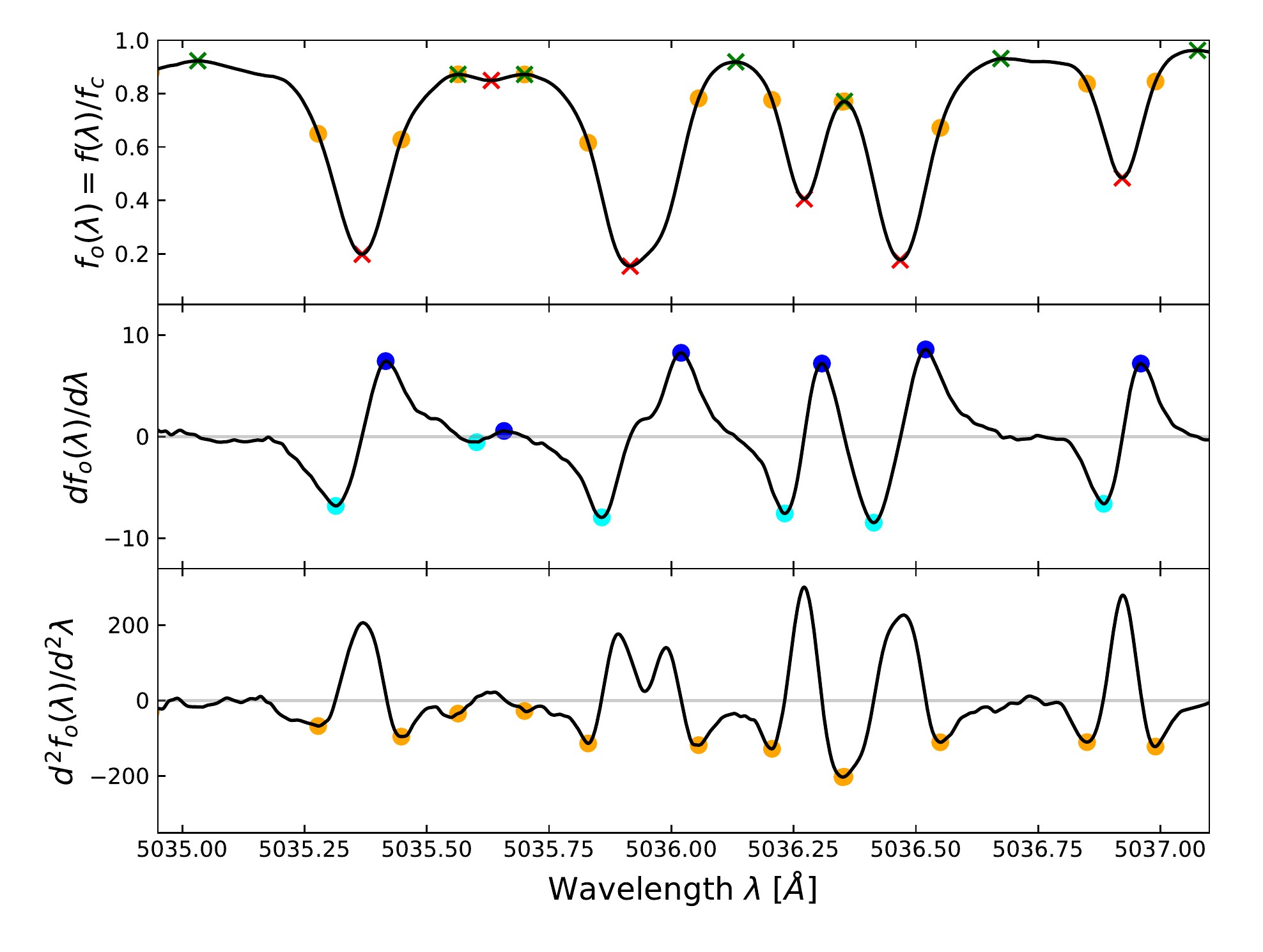}
        \caption{Main points for deriving our morphological parameters. Only the 5035.9 $\AA$ line is blended in this case. \textbf{First row}: Master quiet spectrum (\textit{black curve}) was used to derive the line centre (\textit{red crosses})  as well as the large window (LW) based on the two extrema (\textit{green crosses}). The points for measuring the asymmetry from the second derivative (\textit{bottom panel}) are also indicated (\textit{orange points}). The minimum distance between them and the line centre defines the small window (SW). \textbf{Second row}: First derivative of the flux. The extrema of the derivative are registered on each side of the lines (\textit{blue points}). \textbf{Last row}: Second derivative of the flux. The first negative minima on each side of the lines are registered.}
        \label{FigSchema2}
\end{figure}

Our morphological parameters are based on a few critical points of the line profile, either computed on the spectrum, or on its first or second derivative, as shown in Fig.~\ref{FigSchema2}.
We note that due to the increase of the noise when we computed the first and second derivative, the obtained series were smoothed using a moving average. In the following description, each important point in the different line profiles shown in Fig.~\ref{FigSchema2} is referred
to using a capital letter corresponding to the colour of these points for simplicity.

The first important points in the spectrum, shown in the top panel of Fig.~\ref{FigSchema2}, are the local minima (red crosses, $R$ points) defining our line centres and the local maxima (green crosses, $G$ points). These local maxima can be used to define the large window (LW$_i$=min($\left| G_i-R_i \right|$)) of the spectral line $i$, which is the maximum symmetric half-window size allowed for a spectral line $i$ when we do not wish to probe neighbouring lines. A first parameter, called \textit{\textup{continuum difference,}} is computed as the difference of the local maxima. This parameter is 0 when the line is symmetric. A second parameter, called \textit{\textup{continuum average,}} is the average of the local maxima and provides a measure of the mean pseudo-continuum surrounding a spectral line. This parameter helps probing lines that are surrounded by equivalent blends. Such a line might still be symmetric, but might present some spurious RV variation induced by change in depth or width of the surrounding blends. Finally, a bisector is computed for each line and the number of extrema visible in the bisector is recorded.

In the first derivative of the spectrum, shown in the middle panel of Fig.~\ref{FigSchema2}, we search for the closest extremum around each line centre (blue points, $B_+$ and $B_-$) and calculate the
centre of mass, $B_{cm}$, of these two points. This should be null with the coordinates $(x,y)=(R_i,0)$ if the line is perfectly symmetric. This $B_{cm}$ value has to be normalised in order to account for the fact that stronger or thinner lines more easily produce a displacement of the centre of mass. This normalisation is performed by dividing the $x$ and $y$ coordinates of $B_{cm}$ by the $x$ and $y$ distances between $B_+$ and $B_-$, respectively.

Finally, in the second derivative of the spectrum, shown in the bottom panel of Fig.~\ref{FigSchema2}, we search for the negative minima around a line centre (orange points, $O$ points) because a symmetric line
should exhibit two of them. These points are found farther out from the line centre than the $B$ points, but closer than the local maxima $G$. We can therefore use them to define the small window (SW$_i$=min($\left|O_i-R_i \right|$)) of the spectral line $i$, which is smaller than LW$_i$. In the difference in flux in the spectrum between these two local minima, we observe high values when the spectral lines are blended, as for the line at 5036 $\AA$ in Fig.~\ref{FigSchema2}. Our fourth parameter, named \textit{\textup{jerk distance}}, is defined as this difference normalised by the line depth, defined as the maximum flux difference between the local maxima of a line and its local minima. Finally, we also count for each line the number of local maxima in the second derivative between the two corresponding $O$ points.

We show in Table~\ref{TableClass2} the criteria we used for the different morphological parameters to distinguish between symmetric and asymmetric spectral lines.
With these criteria for the small part of the spectrum shown in Fig.~\ref{FigSchema2}, only the 5035.9 $\AA$ line is not classified as symmetric.

\renewcommand{\arraystretch}{1.6}
\begin{table}[ht]
        \caption{Criterion for selecting the symmetric spectral lines. The reference values for a perfect symmetric line are listed in the last column. }
        \label{TableClass2}
        \centering
        \begin{tabular}{ccc}
                \hline\hline

                Category & Criterion & Symmetric \\
                \hline

                Continuum difference & $<0.30$ & 0\\
                Continuum average & $>0.80$ & 1\\
                SW & $>0.04$ & -\\
                Mass centre (derivative) & $<0.20$ & 0\\
                Jerk distance & $<0.25$ & 0\\
                Number of bisector extrema & $<4$ & $\lesssim 2$\\
                Number of maxima (second derivative)  & $1$ & 1  \\
                \hline
        \end{tabular}
\end{table}

\section{Simulation of the stellar activity effect on the convective blueshift inhibition}
\label{appendixC}

In this section, we investigate whether the curve we obtained in Fig.~\ref{FigReiners3} is consistent with the inhibition of the convective blueshift observed in the Sun.
\citet{Cavallini(1985)} studied the variation in the bisector for three iron spectral lines when observed from a quiet photospheric solar region to the centre of a faculae,
and reported this for different centre-to-limb angles $\mu$. By definition, $\mu=\cos{\theta}$, where $\theta$ is the angle between us, the solar centre, and a given position on the
solar disc. $\mu$ is therefore zero at disc centre, and one in the extreme part of the limb.

We digitalised Figure 1, 5, 6, and 7 in \citet{Cavallini(1985)} and studied the bisectors $b$ of the quiet solar photosphere and at the centre of the active regions for different $\mu$ angles.
We then fitted a 2D polynomial, with a maximum of five orders in flux and six in $\mu$, to model the variations in the quiet and active bisector as a function of $\mu,$ 
\begin{equation}\label{eq:bis}
b [km/s] = \sum_{i=0}^6 \sum_{j=0,i+j\leq6}^5 C_{i,j} \cdot \mu^i f^{j}.
\end{equation}
The coefficients for the polynomial fit of the quiet and the active bisectors for the 6297.8, 6301.5, and 6302.5\,$\AA$ spectral lines can be found in Table~\ref{TableCoeff1}.  \citet{Cavallini(1985)} also reported that when $\mu$ increases, the depth of the bisector and therefore the line depth decreases. We also modelled this effect using a simple first-order polynomial in $\mu$, where $f_0$ if the flux in the line core at disc centre,
\begin{equation}\label{eq:bis_depth}
f_{min} = f_0 + \beta (1-\mu).
\end{equation}
The corresponding coefficients can be found at the bottom of Table~\ref{TableCoeff1}.

\renewcommand{\arraystretch}{1.6}
\begin{table*}[ht]
        \caption{Coefficients $C_{i,j}$ shown in Eq.~\ref{eq:bis} that we used to model the bisectors of the three iron lines at 6297.8, 6301.5, and 6302.5\,$\AA$ in the quiet photosphere and in a facula region. At the bottom of the table, we display the coefficients $f_0$ and $\beta$ that appear in Eq.~\ref{eq:bis_depth} and were used to model the weakening of the lines away from disc centre.}
        \label{TableCoeff1}
        \centering
        \begin{tabular}{c|ccc|ccc}
                \hline\hline
                 &  \multicolumn{3}{c}{Quiet photosphere} & \multicolumn{3}{c}{Faculae}\\

                Terms & 6297.8 \AA  & 6301.5 \AA & 6302.5 \AA  & 6297.8 \AA  & 6301.5 \AA & 6302.5 \AA \\
                \hline

                constant&-7.77&-6.54&-0.60 &-2.03&-9.49&-18.11\\
                $\mu$ & 3.35&4.51&-7.15 &30.41&18.15&43.42\\
                $\mu^2$ & 8.62&39.93&29.13&-91.58&-41.86&-67.89\\
                $\mu^3$ &-3.74&-72.46&-48.87&116.11&64.95&136.39\\
                $\mu^4$ &-22.77&37.67&28.43&-114.79&-65.84&-180.56\\
                $\mu^5$ &29.82&5.40&1.42&100.20&48.34&136.06\\
                $\mu^6$ &-10.28&-7.06&-4.61&-39.97&-16.57&-43.81\\
                $f$ & 68.09&60.37&8.61&14.35&84.10&137.63\\
                $f \cdot \mu$ & -43.42&-89.10&25.23&-146.89&-113.71&-277.55\\
                $f \cdot \mu^2$ & -40.46&-93.56&-78.08&379.01&173.26&221.14\\
                $f \cdot \mu^3$ &86.84&225.83&137.23&-309.45&-180.82&-231.28\\
                $f \cdot \mu^4$ &-50.26&-146.86&-93.92&78.95&72.43&140.08\\
                $f \cdot \mu^5$ &6.02&32.37&23.01&1.81&-8.09&-29.49\\
                $f^2$ &-211.77&-195.13&-21.91&-27.87&-276.11&-402.35\\
                $f^2\cdot \mu$ &138.67&339.15&-63.71&220.83&264.86&713.25\\
                $f^2 \cdot \mu^2$ &9.70&-39.32&46.47&-557.60&-247.76&-304.21\\
                $f^2 \cdot \mu^3$ &-61.95&-108.82&-63.23&367.89&200.37&171.68\\
                $f^2 \cdot \mu^4$ &29.34&42.38&21.94&-68.81&-49.55&-56.92\\
                $f^3 $ &320.97&302.32&24.69&23.06&447.00&583.29\\
                $f^3 \cdot \mu$ &-188.31&-513.81&114.60&-91.70&-311.38&-936.23\\
                $f^3 \cdot \mu^2$ &33.14&135.83&-6.77&304.97&125.34&196.17\\
                $f^3 \cdot \mu^3$ &-1.04&6.31&7.85&-120.12&-57.90&-33.64\\
                $f^4$ &-239.55&-225.28&-11.70&-7.51&-358.94&-419.98\\
                $f^4 \cdot \mu$ &111.87&343.85&-109.78&-47.17&192.28&621.77\\
                $f^4\cdot \mu^2$ &-10.96&-58.89&0.41&-45.97&-15.88&-53.15\\
                $f^5$ &70.77&64.15&1.08&0.78&114.64&120.32\\
                $f^5 \cdot \mu$ &-24.54&-83.02&40.58 &33.76&-50.79&-165.70\\
                \hline\hline

        $f_0$&0.382&0.296&0.374 &0.411&0.350&0.434\\
        $\beta$ & 0.129&0.143&0.153 &0.149&0.078&0.141\\
                \hline
        \end{tabular}
\end{table*}

We then implemented these models in the SOAP\,2.0 code \citep[][]{Dumusque(2014)} to estimate the effect induced by the inhibition of convection inside a faculae. SOAP\,2.0 requires as input
a line profile for the quiet photosphere and the active region we wish to model. We injected in both cases a symmetric line profile, and then used our models of the bisector shape and bisector depth to estimate the shape of the 6297.8, 6301.5, or 6302.5\,$\AA$ spectral lines throughout the surface of the simulated star in SOAP\,2.0. To obtain a realistic symmetric line profile, we fitted a Voigt profile to the 6297.8, 6301.5, and 6302.5\,$\AA$ spectral lines in the Kitt Peak solar atlas obtained at disc centre \citep[][]{Wallace-1998}, and used these profiles as input for SOAP\,2.0.

As shown in \citet{Gray(2009)}, when spectral lines formed by Fe I 
are considered, each line is affected by the same C-shaped bisector, truncated at the line depth. Following this observation, we
used the same models fitted on the 6297.8, 6301.5, and 6302.5\,$\AA$ spectral lines to model shallower spectral lines.

We initiated a SOAP\,2.0 simulation to the solar case \citep[T$_{\mathrm{eff}}=5778\,K$, P$_{\mathrm{rotation}}=25$ days, inclination=90 degrees, quadratic limb darkening=(0.29,0.34),][]{Oshagh-2013a}) with a facula of size 1\% on the equator. We then ran several simulations for the three different Fe I 
spectral lines and for different depths from 5 to 100\% of the original line depths, and recorded for each simulation the semi-amplitude of the activity signal modelled with SOAP\,2.0. The results of these simulations are shown in Fig~\ref{FigSimu}.

Fig~\ref{FigSimu} shows that when the values of the y-axis are ignored, the SOAP\,2.0 simulation for the spectral lines at 6297.8 and 6301.5\,$\AA$ show a similar behaviour as in Fig.~\ref{FigReiners3} when we consider the amplitude of the activity signal as a function of line depth. At shallow spectral lines, the semi-amplitude of the RV activity signal follows an exponential growth, and for lines stronger than 0.5, the RV activity signal reaches a floor. The line at 6302.5\,$\AA$ also shows the increase in the semi-amplitude of the RV activity signal at shallow spectral lines, but an increase is also observed towards strong spectral lines. This can be explained by the bottom part of the bisector of this line, which shows a significant redshift in \citet{Cavallini(1985)} compared to the 6297.8 and 6301.5\,$\AA$ lines. Because these lines all have a similar depth, this difference in bisector shape would imply that the convection close to the surface, where the line core is formed, affects the 6302.5\,$\AA$ spectral line differently, which is unphysical. Bisector computation is very sensitive to blends and telluric line contamination, which are common in this red part of the visible, therefore it is possible that the bottom part of the bisector of the 6302.5\,$\AA$ line is contaminated.

In conclusion, the SOAP\,2.0 simulation performed here shows that the dependence of the semi-amplitude of the RV activity signal as a function of line depth observed in Fig.~\ref{FigReiners3} can be reproduced from
solar observations, at least for the Fe I 
spectral lines at 6297.8 and 6301.5\,$\AA$. Of course, the exact comparison is difficult because the result shown in Fig.~\ref{FigReiners3} is obtained for $\alpha$ Cen B and the simulation here
is based on solar observations. A more detailed comparison will be possible when we analyse the spectra and RVs obtained by the HARPS-N solar telescope \citep[][]{Collier-Cameron:2019aa, Dumusque(2015)}

\end{appendix}

\end{document}